\documentclass[11pt]{iopart}

\usepackage{iopams}
\usepackage{graphicx}
\usepackage{amssymb}
\usepackage{amsfonts}
\usepackage{hyperref}
\usepackage{xcolor}
\usepackage{iopams}
\usepackage[textwidth=17cm]{geometry}
\makeatletter
\let\@fnsymbol\@arabic
\makeatother
\begin{document}
\title{Optical response of a dual membrane active-passive optomechanical cavity}
\author{Akash Kundu$^{1,*}$ \& Chao Jin$^{2}$ \& Jia-Xin Peng$^{3,\dagger}$}
\address{$^1$Institute of Theoretical and Applied Informatics, Polish Academy of Sciences, Baltycka 5, 44-100 Gliwice, Poland}
\address{$^2$Institute for Computational Materials Science, School of Physics and electronics, International Joint Research Laboratory of New Energy Materials and Devices of Henan Province, Henan University, Kaifeng 475004, People’s Republic of China.}
\address{$^3$State Key Laboratory of Precision Spectroscopy, Quantum Institute for Light and Atoms, Department of Physics, East China Normal University, Shanghai 200062, China}
\ead{kunduaku339@gmail.com$^*$}
\ead{18217696127@163.com$^\dagger$}
\vspace{10pt}
\begin{abstract}
{
	We investigate a dual membrane active-passive cavity where each
	mechanical membrane individually quadratically coupled to passive and active cavities
	via two-phonon process. Due to the fact that in the quadratically coupled
	optomechanical system mean-field approximation fails, hence to analyze the system
	completely, we switch to a more generalized out of equilibrium approach, namely
	Keldysh Green’s functional approach. We calculate transmission rate using predetermined
	full retarded Green’s function, and then numerically examine the effect of
	the various parameters on the transmission coefficient and discuss the features and
	physics behind them in detail. On the basis of the optical responsivity we further extend
	our study of fast and slow light phenomenon. The results show that our proposed system
	can not only realize ultra-fast light/ultra-slow light under proper choice of cavity
	parameters, but realization of the conversion between fast and slow light and vice versa.
}
\end{abstract}
\vspace{10pt}
(\small Some figures may appear in colour only in the online journal)
\section{Introduction}
Optomechanical cavities where light coupled to mechanical motion by radiation pressure\cite{0} is a well-developed field for researchers to observe phenomena like gravitational wave detection\cite{1,2}, laser cooling\cite{3,4,5}. A widely used cavity optomechanical system is represented by a single mode Febry-Perot cavity with one movable end, other type of cavities such as Laguerre-Gaussian rotational cavity\cite{5.2,5.4},  rovibrational cavity\cite{5.1,5.3}, active-passive cavity\cite{5.5} are also widely used to set up optomechanical systems. The interaction between cavity and mechanical degree of freedom arises due to the circulating radiation pressure of light intensity. Linear and nonlinear interaction between photon and phonon in an optomechanical system offers an interface for carrying out many interesting phenomena listed as follows. For linear interaction, we can observe ground state cooling of mechanical mirror\cite{6,7,8,9,10}, electromagnetic induced transparency (EIT)\cite{11,12,13,14,14.1}, optomechanical induced transparency (OMIT)\cite{15,16,16.5} (OMIT is formally equivalent to EIT in a cavity optomechanical system operating in the resolved sideband regime\cite{17}), entanglement between light and mirror\cite{18,19,20,20.1,20.2,21,22}, squeezing\cite{24}, normal mode splitting\cite{25,26}, measurement of electric charge\cite{27}. Under consideration of quadratic interaction (where optical cavity mode is coupled to the square of the position of mechanical oscillator) we can observe several interesting phenomena similar as well as distinct from linear interaction. Such as, weak force measurement\cite{28,29}, cooling of mechanical oscillator\cite{30,31,32}, phonon squeezing\cite{23,23.5}, strong photon correlation\cite{33,34,35}, tunable slow light\cite{36}, photon blockade\cite{37}, OMIT\cite{12,37.1}. Here we point out in case of linear coupling (i.e. single-phonon process) mean displacement of an oscillator is nonzero which leads to the modulation of the output lasing field, on the other side
for quadratic coupling (i.e. for two-phonon processes) the mean response of the output
oscillator is null\cite{null}. Hence, modulation of output fields must arise from the mean values
of the square of displacement fluctuations of the mirrors in the quadratic coupling system, which is temperature
dependent. In the following article more attention has been paid to the quadratic optomechanical coupling system as it is worth exploring multi-mode macroscopic quantum phenomena in such system.\\

Electromagnetic Induced Transparency (EIT)\cite{eit1,eit2,eit3,eit4} is a quantum inteference phenomena which usually appears in multi-level atomic system driven by strong pump fields. Since the time it was first theoretically predicted and experimentally verified, EIT has been extensively studied. The studies on EIT helped revealing lot of interesting phenomena such as enhanced nonlinearity\cite{enhanced_nonlinearity}, quantum memories for photons\cite{memory1,memory2}, slow light\cite{slow1,slow2} etc. Optomechanical Induced Transparency (OMIT) is a phenomena analogous to EIT was first investigated in vibrational optomechanical cavity system\cite{15}, whose fundamental mechanism is destructive interference between different pathways of internal fields inside the considered optomechanical system. Recently, OMIT was suggested to
apply in many fields including four-wave mixing\cite{omit1}, quantum router\cite{omit2}, and
precision measurement\cite{5.2,omit3,omit4}, etc. In EIT and OMIT the tunability of slow and fast light effects with the help of strong pump field is found out to be an interesting realization. The group velocity of light is different from that in vacuum is a feature of fast and slow light, which have attracted considerable attention and made remarkable progress\cite{group1,group2,group3} over the past decades because of its potential applications in optical storage\cite{optical_storage}, on-chip optical signal processing\cite{quantum_processing} as well as optical fiber communication systems and networks\cite{group1}.\\

Keldysh functional integral can be derived by direct functional integral quantization of Markovian quantum masters equation\cite{keldysh1}. While dealing with many-body systems an investigation using Keldysh approach is recommended because an externally driven open many-body system can be excellently described by microscopic Markovian master equation, but masters equation can not be efficiently used in a system where the characteristic complications (such as domination of nonlinear interaction, unavoidable quantum fluctuations) of many-body
system are considered so, in the framework of equilibrium many-body physics the solution of these system are impossible to achieve.\\
Due to the appearance of \textit{forward} and \textit{backward} branches of time evolution, Keldysh functional integral is close to real time i.e. von Neumann evolution of quantum mechanics\cite{contour}. To get an analytical description of many-body systems we need express the system's partition function in terms of Keldysh action ($S$) and Keldysh classical ($\psi_{cl}$) and quantum ($\psi_{qu}$) rotations. A variation of partition function with source terms help us to produce two basic types of observables in many-body systems: correlation and
response functions, in terms of retarded ($G^{R}$), advanced ($G^A$), and Keldysh ($G^K$) Green’s functions\cite{keldysh1,keldysh2,keldysh3,keldysh4,keldysh5,keldysh6,keldysh7}. $G^{R/A}$ describes the linear response of a system which is perturbed by a weak external source field, on the other hand $G^K$ efficiently carry
elementary information on the system’s correlations and
occupation of individual quantum mechanical modes. Moreover Keldysh Green's functional approach treats spatial and temporal correlations in a many-body system on equal footing unlike master equation where the spatial correlation is accessible but not the temporal one (which need to be calculated using quantum regression theorem\cite{opensystem_book,57}).\\

Based on the previous discussion, in this work, we theoretically studied the optical
response of a dual membrane active-passive optomechanical cavity, where each
mechanical membrane individually quadratically coupled to passive and active cavities
via two-phonon process. It needs to be pointed out that: in case of quadratic coupling
(i.e. for two-phonon processes) the mean response of the mechanical membrane is null, which lead to the invalidation of mean-field approximation. So, instead of mean-field
theory approach we used more analytical Keldysh functional approach related to Green’s functions,
which allows the quantum fluctuation effect of the system to be taken into account.\\

We have mainly investigated the system under following three assumptions. (1)
There is only one mechanical membrane, (2) there are two identical mechanical
membranes, and (3) there are two distinct mechanical membranes inside the active-passive
cavity optomechanical system. Our results show that the appeared number of
transparency windows depend on how many distinct effective mechanical frequencies
in the system. In particular, we find that, with the increase of the effective pump rate of
active cavity, the transparent window caused by the two-phonon process gradually
becomes shallower and then converted into an absorption peak, and the absorption
coefficient can be up to 1 or more. On the basis of the optical responsivity we further
extend our study of fast and slow light phenomenon. The results show that our proposed
system can not only realize ultra-fast light/ultra-slow light under proper choice of cavity
parameters, but also achieve the conversion of fast and slow light by adjusting the
effective pump rate of active cavity or the photon-tunneling strength.\\

{The article organized as follows, in Section \ref{sec:sec1}, we obtain the effective model by defining Hamiltonian of the system under Keldysh action framework.
In Section \ref{sec:sec2}, we study the influence of the various parameters on the optical responsivity
of the system and discuss physics behind them in detail. On the basis of the observation made in Section \ref{sec:sec2} we have extended our results to investigate fast and slow light effects by changing optical group delay $\tau_{g}$ with several parameters of the system in Section \ref{sec:sec3}. The article has been summerized and concluded in Section \ref{sec:sec4}.}
\section{The Model and Hamiltonian}\label{sec:sec1}
\begin{figure}
	\centering\includegraphics[width=0.9\linewidth]{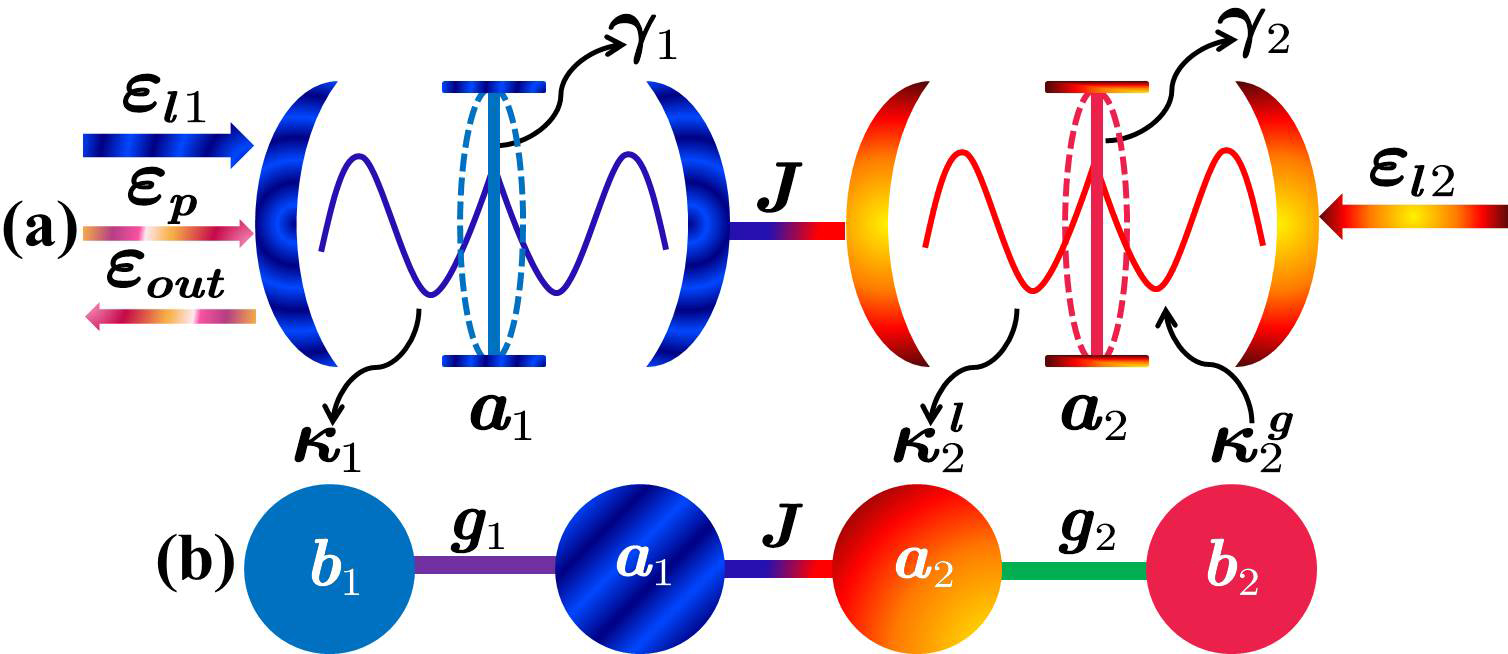}
	\caption{\small (a) Diagrammatic representation of the active(right)-passive(left) cavity optomechanical system, here $J$ is the photon hopping rate between the two cavities. The passive (active) cavity mode $\hat{a}_1$($\hat{a}_2$) is quadratically coupled to the displacement of membrane. $\kappa_{1}$($\kappa_{2}^l$) denote the loss rate of passive (active) cavity and $\gamma_{1}$ ($\gamma_{2}$) is decay rate of the mechanical membrane. The active cavity is strongly driven by incoherent pumping with the pump rate $\kappa_{2}^g$. The passive (active) cavity is optically driven by a strong control field defined by strength $\epsilon_{l1}$ ($\epsilon_{l2}$). A weak field with amplitude $\varepsilon_p$ is used to probe the passive cavity, and $\varepsilon_{out}$ is the amplitude of output field. (b) The diagram of interactions among subsystems in the double optomechanical cavity.}
	\label{fig:fig1}
\end{figure}

The system we propose is sketched in Fig.\ref{fig:fig1}. It consists of a passive cavity and an active one, where a thin dielectric membrane is placed inside each cavity. When the membrane is located in the node (or antinode) of the cavity field, the cavity frequency will quadratically couple to the membrane displacement. The system non-Hermitian Hamiltonian under these considerations (including gain and dissipation) can be written as $\left(\hbar =1\right)$
\begin{eqnarray}
H_{1non}&=&\left(\omega _{a}-\frac{\kappa _{1}}{2}i\right) \hat{a}_{1}^{\dagger
}\hat{a}_{1}+\left( \omega _{a}+\frac{\kappa _{2}}{2}i\right) \hat{a}_{2}^{\dagger
}\hat{a}_{2}-J\left(\hat{a}_{1}^{\dagger }\hat{a}_{2}+\hat{a}_{2}^{\dagger }\hat{a}_{1}\right)+\left(\omega_{m1}-\frac{\gamma _{1}}{2}i\right)\hat{b}_{1}^{\dagger
}\hat{b}_{1} \nonumber\\
&&+\left( \omega _{m2}-\frac{\gamma_{2}}{2}i\right) \hat{b}_{2}^{\dagger}\hat{b}_{2}+g_{1}\hat{a}_{1}^{\dagger }\hat{a}_{1}\left( \hat{b}_{1}^{\dagger }+\hat{b}_{1}\right) ^{2}+g_{2}\hat{a}_{2}^{\dagger}\hat{a}_{2}\left( \hat{b}_{2}^{\dagger }+\hat{b}_{2}\right) ^{2} \nonumber\\
&&+i\left[ \left( \varepsilon _{l1}e^{-i\omega_{l}t}+\varepsilon
_{p}e^{-i\omega_{p}t}\right)\hat{a}_{1}^{\dagger }-\mathrm{H.c.}\right]+i\varepsilon _{l2}\left(\hat{a}_{2}^{\dagger }e^{-i\omega_{l}t}-\hat{a}_{2}e^{i\omega_{l}t}\right)\label{eq:non-Hermitian-Hamiltonian1},
\end{eqnarray}
where the first and two terms describe the energy of the passive cavity and active one, respectively, with same resonance frequency $\omega_{a}$. The third term stands for  coupling between the passive cavity with active one through hopping rate $J$.
The fourth and fifth terms represent the energy of the membrane in the two cavities with resonance frequency $\omega_{m1}$ and $\omega_{m2}$, respectively. The sixth and seventh terms are the quadratic interaction between passive and active cavity modes with moving membranes through optomechanical coupling constant $g_1$ and $g_2$, respectively. The last two terms represent the interaction between the cavity field with the pumping field and the weak probe field, with driving strength $\varepsilon_{l1}$, $\varepsilon_{l2}$ (for strong field) and $\varepsilon_p$ (for weak field). In Eq.\ref{eq:non-Hermitian-Hamiltonian1}, $\hat{a}_1$($\hat{a}_1^\dagger$) and $\hat{a}_2$($\hat{a}_2^\dagger$) are annihilation (creation) operators of the passive cavity and active one, respectively; $\hat{b}_1$($\hat{b}_1^\dagger$) and $\hat{b}_2$($\hat{b}_2^\dagger$) are the annihilation (creation) operators of the membranes situated inside passive and active cavity respectively; $\kappa_{1}=\kappa_{1}^0+\kappa_{1}^e$ is the total loss rate of passive cavity, here $\kappa_{1}^0$ and $\kappa_{1}^e$ are the intrinsic and external loss rate, respectively; $\kappa_{2}=\kappa_{2}^g-\kappa_{2}^l$ is the effective pump rate of active cavity; $\gamma_{1}$ ($\gamma_{2}$) is decay rate of the mechanical membrane. The above Hamiltonian can be rewritten in a frame rotating at frequency $\omega _{l}$ as
 \begin{eqnarray}
 H_{2non}&=&\left(\Delta _{a}-\frac{\kappa _{1}}{2}i\right) \hat{a}_{1}^{\dagger
 }\hat{a}_{1}+\left(\Delta _{a}+\frac{\kappa _{2}}{2}i\right)\hat{a}_{2}^{\dagger
 }\hat{a}_{2}-J\left(\hat{a}_{1}^{\dagger }\hat{a}_{2}+\hat{a}_{2}^{\dagger }\hat{a}_{1}\right)+\left(\omega _{m1}-\frac{\gamma _{1}}{2}i\right)\hat{b}_{1}^{\dagger
 }\hat{b}_{1}\nonumber\\
 &&+\left(\omega _{m2}-\frac{\gamma _{2}}{2}i\right)\hat{b}_{2}^{\dagger
 }\hat{b}_{2}+g_{1}\hat{a}_{1}^{\dagger }\hat{a}_{1}\left(\hat{b}_{1}^{\dagger }+\hat{b}_{1}\right) ^{2}+g_{2}\hat{a}_{2}^{\dagger}\hat{a}_{2}\left( \hat{b}_{2}^{\dagger}+\hat{b}_{2}\right) ^{2}\nonumber\\
 &&+i\left[ \left(\varepsilon _{l1}+\varepsilon_{p}e^{-i\delta t}\right)
 \hat{a}_{1}^{\dagger}-\mathrm{H.c}\right] +i\varepsilon _{l2}\left(\hat{a}_{2}^{\dagger }-\hat{a}_{2}\right),\label{eq:modified-hamiltonian}
 \end{eqnarray}
where $\Delta _{a}=\omega _{a}-\omega _{l}$ is the detuning of the cavity mode from the pumping field, and  $\delta =\omega _{p}-\omega _{l}$ is the detuning of the probe field from the pumping field.

\subsection{Steady State Approximation}
In order to study the mean response of the active-passive cavity system to the weak probe field in the presence of the pumping field, we use the Heisenberg equations, we can obtain the mean-value equations of the system operators as follows 
	\begin{equation}
	\frac{\textrm{d}\left\langle \hat{a}_{1}\right\rangle }{dt}=-i\left[ \Delta
	_{a}+g_{1}\left\langle \left(\hat{b}_{1}^{\dagger }+\hat{b}_{1}\right)
	^{2}\right\rangle \right] \left\langle \hat{a}_{1}\right\rangle -\frac{\kappa_{1}%
	}{2}\left\langle\hat{a}_{1}\right\rangle
	+iJ\left\langle\hat{a}_{1}\right\rangle
	+\varepsilon _{l1},
	\end{equation}
	\begin{equation}
	\frac{\textrm{d}\left\langle\hat{a}_{2}\right\rangle}{dt}=-i\left[\Delta
	_{a}+g_{2}\left\langle \left(\hat{b}_{2}^{\dagger }+\hat{b}_{2}\right)
	^{2}\right\rangle\right]\langle \hat{a}_{2}\rangle+\frac{\kappa _{2}%
	}{2}\langle\hat{a}_{2}\rangle
	+iJ\langle\hat{a}_{1}\rangle
	+\varepsilon_{l2},
	\end{equation}
	\begin{equation}
	\frac{\textrm{d}\langle \hat{b}_{1}\rangle}{dt}=-i\left(\omega
	_{m1}+2g_{1}\langle\hat{a}_{1}^{\dagger }\rangle\langle
	\hat{a}_{1}\rangle\right)\langle \hat{b}_{1}\rangle-\frac{\gamma _{1}}{2}%
	\langle\hat{b}_{1}\rangle -2ig_{1}\langle\hat{a}_{1}^{\dagger
	}\rangle\langle\hat{a}_{1}\rangle\langle \hat{b}_{1}^{\dagger
	}\rangle,
	\end{equation}
	\begin{equation}
	\frac{\textrm{d}\langle\hat{b}_{2}\rangle}{dt}=-i\left(\omega_{m2}+2g_{2}\langle\hat{a}_{2}^{\dagger}\rangle\langle\hat{a}_{2}\rangle\right)\langle\hat{b}_{2}\rangle-\frac{\gamma_{2}}{2}\langle\hat{b}_{2}\rangle-2ig_{2}\langle\hat{a}_{2}^{\dagger
	}\rangle\langle\hat{a}_{2}\rangle \langle\hat{b}_{2}^{\dagger
	}\rangle.
	\end{equation}
Note that, in the above equation we have assumed input noises of cavity mode and mechanical membrane are zero (i.e. $\langle \hat{a}_{1in}\rangle=\langle \hat{a}_{2in}\rangle=0$) and $\langle \hat{b}_{1in}\rangle=\langle \hat{b}_{2in}\rangle=0$) and adopted the mean field approximation \cite{56.5} under the strong pumping field regime, viz,
\begin{eqnarray}
\qquad\qquad\qquad\qquad\qquad&&\langle\hat{a}_1^\dagger\hat{a}_1\rangle\approx\langle\hat{a}_1^\dagger\rangle\langle\hat{a}_1\rangle\nonumber,\\&&\\
&&\langle\hat{a}_2^\dagger\hat{a}_2\rangle\approx\langle\hat{a}_2^\dagger\rangle\langle\hat{a}_2\rangle\nonumber.
\end{eqnarray}
Therefore, we are working in a classical regime and do not include quantum fluctuation effects. According to the mean-value equations, setting $\langle\hat{\dot{a}}_1\rangle=\langle\hat{\dot{a}}_2\rangle=\langle\hat{\dot{b}}_1\rangle=\langle\hat{\dot{b}}_2\rangle=0$, the steady-state values of the dynamical variables are easily found to be 
	\begin{eqnarray}
	\qquad\qquad\qquad\qquad&&\left\langle\hat{a}_{1}\right\rangle=\frac{iJ\varepsilon _{l2}+\varepsilon
		_{l1}\left( i\Delta _{a}-\frac{\kappa _{2}}{2}\right) }{\left( i\Delta _{a}+%
		\frac{\kappa _{1}}{2}\right) \left( i\Delta _{a}-\frac{\kappa _{2}}{2}%
		\right) +J^{2}}, \\
	&&\left\langle\hat{a}_{2}\right\rangle =\frac{iJ\left\langle\hat{a}_{2}\right\rangle+\varepsilon_{l2}}{i\Delta _{a}-\frac{\kappa _{2}}{2}}, \\
	&&\langle\hat{b}_{1}\rangle =\langle\hat{b}_{2}\rangle=0.
	\end{eqnarray}
At steady state, $\langle\hat{b}_1\rangle=\langle\hat{b}_2\rangle=0$ indicates that the amplitude of cavity field is unrelated to the displacement of mechanical membrane, hence the output field is not modified by mean displacement of the mechanical membrane, which lead to mean field approximation fails. 
\subsection{Quantum Fluctuations Effect}
Next, we consider the quantum fluctuation effect on the active-passive cavity system. Under strong driving field, the cavity mode can be split into a steady-state value and a fluctuation term in the usual way as
\begin{eqnarray}
\qquad\qquad\qquad\qquad\qquad&&\hat{a}_1=\langle\hat{a}_1\rangle+\hat{A}_1\nonumber,\\&\label{eq:fluctuation_approx}&\\
&&\hat{a}_2=\langle\hat{a}_2\rangle+\hat{A}_2\nonumber,
\end{eqnarray}
without loss of generality the $\langle\hat{a}_1\rangle, \langle\hat{a}_2\rangle$ can be chosen as a real. Substituting Eqs.\ref{eq:modified-hamiltonian} into Eq.\ref{eq:fluctuation_approx}, we can obtain the following linearized effective Hamiltonian
\begin{eqnarray}
H_{eff}&=&\left(\Delta _{a}-\frac{\kappa _{1}}{2}i\right)\hat{A}_{1}^{\dagger
}\hat{A}_{1}+\left(\Delta_{a}+\frac{\kappa _{2}}{2}i\right)\hat{A}_{2}^{\dagger
}\hat{A}_{2}-J\left(\hat{a}_{1}^{\dagger }\hat{a}_{2}+\hat{a}_{2}^{\dagger}\hat{a}_{1}\right)-J\left(\hat{A}_{1}^{\dagger}\hat{A}_{2}+\hat{A}_{2}^{\dagger}\hat{A}_{1}\right)\nonumber\\ 
&&+\left(\omega_{b1}-\frac{\gamma _{1}}{2}i\right) \hat{b}_{1}^{\dagger }\hat{b}_{1}+\left( \omega_{b2}-\frac{\gamma _{2}}{2}i\right) \hat{b}_{2}^{\dagger }\hat{b}_{2}+G_{1}(\hat{A}_{1}^{\dagger}\hat{b}_{1}^{2}+\hat{A}_{1}\hat{b}_{1}^{\dagger^{2}})\nonumber\\
&&+G_{2}(\hat{A}_{2}^{\dagger}\hat{b}_{2}^{2}+\hat{A}_{2}\hat{b}_{2}^{\dagger^{2}})\label{eq:linerize-hamiltonian},
\end{eqnarray}
where $\omega _{b1}=\omega _{m1}+2g_{1}\left\vert \left\langle
\hat{a}_{1}\right\rangle \right\vert ^{2},$ $\omega _{b2}=\omega_{m2}+2g_{2}\left\vert \left\langle\hat{a}_{2}\right\rangle\right\vert^{2}$ are the shifted mechanical frequency; $G_{1}=g_{1}\left\langle \hat{a}_{1}\right\rangle,$ $G_{2}=g_{2}\left\langle\hat{a}_{2}\right\rangle$ are the effective optomechanical coupling strength.
It should be noted that in the Eq.\ref{eq:linerize-hamiltonian} we have  omitted fourth-order small terms (i.e. $\hat{A}_{2}^{\dagger^2}\hat{b}_{2}^{2}$) and make a rotating-wave approximation (considering the red detuned sideband, $\omega_{a}-\omega_{l}>0$), that is
\begin{eqnarray}
\qquad\qquad\qquad&&G_{1}\left(\hat{A}_{1}^{\dagger }+\hat{A}_{1}\right)\left(\hat{b}_{1}^{\dagger}+\hat{b}_{1}\right)^{2}\rightarrow G_{1}(\hat{A}_{1}^{\dagger
}\hat{b}_{1}^{2}+\hat{A}_{1}\hat{b}_{1}^{\dagger ^{2}})\nonumber, \\\label{eq:last-two-term}\\
&&G_{2}\left(\hat{A}_{2}^{\dagger }+\hat{A}_{2}\right) \left(\hat{b}_{2}^{\dagger
}+\hat{b}_{2}\right)^{2}\rightarrow G_{2}(\hat{A}_{2}^{\dagger
}\hat{b}_{2}^{2}+\hat{A}_{2}\hat{b}_{2}^{\dagger ^{2}})\nonumber.
\end{eqnarray}
To observe dynamics of the system, we can use the master equation which is given by \cite{57}
\begin{eqnarray}
\frac{\textrm{d}\rho }{\textrm{d}t}&=&-i\left[H_{eff},\rho \right] +\frac{\kappa _{1}}{2}%
\mathcal{L}\left[\hat{A}_{1}\right] \rho +\frac{\kappa_{2}^{l}}{2}\mathcal{L}%
\left[\hat{A}_{2}\right] \rho+\frac{\kappa _{2}^{g}}{2}\mathcal{L}\left[\hat{A}_{2}^{\dagger }\right]\rho+\frac{\gamma_{1}}{2}\left( n_{1th}+1\right)\mathcal{L}\left[\hat{b}_{1}%
\right] \rho\nonumber,\\
&&+\frac{\gamma_{1}}{2}n_{1th}\mathcal{L}\left[\hat{b}_{1}^{\dagger }\right] \rho+\frac{\gamma_{2}}{2}n_{2th}\mathcal{L}\left[\hat{b}_{2}^{\dagger}\right]\rho+\frac{\gamma_{2}}{2}\left(n_{2th}+1\right)\mathcal{L}\left[\hat{b}_{2}\right] \rho
\label{eq:master-equation},
\end{eqnarray}
where $\mathcal{L}\left[\hat{o}\right] \rho =2\hat{o}\rho\hat{o}-\left(\hat{o}^{\dagger}\hat{o}\rho+\rho\hat{o}^{\dagger}\hat{o}\right) $ is the standard dissipator in the Lindblad
form, and $n_{ith}=$\\ 
$\left[ \exp \left( \hbar \omega _{mi}/k_{B}T\right) -1%
\right] ^{-1}\left(i=1,2\right)$ is the average thermal phonon numbers of the mechanical bath at temperature $T$. However, generally speaking, it is very difficult to solve the master equation, which can only be solved numerically. Therefore, to get a fully quantum description, we turn to the formulations under Keldysh framework \cite{keldysh1,58}. At first, we need to write the partition function of system (it is fully equivalent to the master  Eq.(\ref{eq:master-equation})) which is written as 
\begin{eqnarray}
Z=\int D\left( \psi _{cl},\psi _{qu}\right) \exp \left[ iS_{0}\left( \psi
_{cl},\psi _{qu}\right) +iS_{I}\left( \psi _{cl},\psi _{qu}\right) \right]\label{eq:partition-function},
\end{eqnarray}
where $\psi _{cl}$ and $\psi _{qu}$ represent the classical and quantum
field, respectively. In Eq.(\ref{eq:partition-function}) $S_0$ corresponds to linear Keldysh action which includes all the linear and dissipative terms of the master equation, and $S_I$ is the nonlinear Keldysh action includes the interaction term of the master equation. In the frequency space, $S_0$ and $S_I$ can be defined by
\begin{eqnarray}
S_{0}=\frac{1}{2\pi }\int_{-\infty }^{+\infty }\left( \psi _{cl}^{\ast
},\psi _{qu}^{\ast }\right) \left( 
\begin{array}{cc}
0 & D_{4\times 4}^{A}\left( \omega \right) \\ 
D_{4\times 4}^{R}\left( \omega \right) & D_{4\times 4}^{K}\left( \omega
\right)%
\end{array}%
\right) \left( 
\begin{array}{c}
\psi _{cl} \\ 
\psi _{qu}%
\end{array}%
\right) d\omega\label{eq:linear-keldysh-action},
\end{eqnarray}
\begin{equation}
S_{I}=S_{I1}+S_{I2}\label{eq:nonlinear-keldysh-action},
\end{equation}
where $S_{I1}$ and $S_{I2}$
\begin{eqnarray}
S_{I1}=-\frac{G_{1}}{\sqrt{2}}\int_{-\infty }^{+\infty}\left(2A_{1cl}b_{1cl}^{\ast }b_{1qu}^{\ast}+A_{1qu}b_{1cl}^{\ast^{2}}+A_{1qu}b_{1qu}^{\ast^{2}}\right.&&+\left.2A_{1cl}^{\ast }b_{1cl}b_{1qu}\right.\nonumber\\
&&\left.+A_{1qu}^{\ast}b_{1cl}^{^{_{2}}}+A_{1qu}^{\ast}b_{1qu}^{^{_{2}}}\right)\label{eq:S_I1},
\end{eqnarray}
\begin{eqnarray}
S_{I2}=-\frac{G_{2}}{\sqrt{2}}\int_{-\infty }^{+\infty}\left(2A_{2cl}b_{2cl}^{\ast }b_{2qu}^{\ast}+A_{2qu}b_{2cl}^{\ast^{2}}+A_{2qu}b_{2qu}^{\ast ^{2}}\right.&&+\left.2A_{2cl}^{\ast }b_{2cl}b_{2qu}\right.\nonumber\\
&&\left.+A_{2qu}^{\ast
}b_{2cl}^{^{_{2}}}+A_{2qu}^{\ast }b_{2qu}^{^{_{2}}}\right)\label{eq:S_I2}.
\end{eqnarray} 
In (\ref{eq:linear-keldysh-action}), $\left( \psi _{cl},\psi _{qu}\right) ^{T}$=$\left(
A_{1cl},A_{2cl},b_{1cl},b_{2cl},A_{1qu},A_{2qu},b_{1qu},b_{2qu}\right) ^{T}, 
$ in which\\
 $A_{icl}=\left( A_{i+}+A_{i-}\right) /\sqrt{2}A_{iqu}=\left(A_{i+}-A_{i-}\right) /\sqrt{2},$ $b_{icl}=\left( b_{j+}+b_{j-}\right) /\sqrt{2},b_{iqu}=\left( b_{j+}-b_{j-}\right) /\sqrt{2}\left( i=1,2\right).$ The terms in $4\times4$ matrix in Eq.(\ref{eq:linear-keldysh-action}) are defined as 
\begin{eqnarray}
\left[ D_{4\times 4}^{R(A)}\right] \left( \omega \right)=\left[ 
\begin{array}{cccc}
\omega -\Delta _{a}\pm \frac{\kappa _{1}}{2}i & -J & 0 & 0 \\ 
-J & \omega -\Delta _{a}\mp \frac{\kappa _{2}}{2}i & 0 & 0 \\ 
0 & 0 & \omega -\omega _{b1}\pm \frac{\gamma _{1}}{2}i & 0 \\ 
0 & 0 & 0 & \omega -\omega _{b2}\pm \frac{\gamma _{2}}{2}i%
\end{array}%
\right],\nonumber\\
\label{mat:keldysh-g1}
\end{eqnarray}
	
	\begin{equation}
	D_{4\times 4}^{K}\left( \omega \right) =\left[ 
	\begin{array}{cccc}
	i\kappa _{1} & 0 & 0 & 0 \\ 
	0 & i\left( \kappa _{2}^{l}+\kappa _{2}^{g}\right) & 0 & 0 \\ 
	0 & 0 & i\gamma _{1}\left( 2n_{1th}+1\right) & 0 \\ 
	0 & 0 & 0 & i\gamma _{2}\left( 2n_{2th}+1\right)%
	\end{array}%
	\right]\label{mat:keldysh-g2}.
	\end{equation}
They are the inverses of the unperturbed retard Green's function($G^R_{4\times4}$), advanced
Green's function($G^A_{4\times4}$), and Keldysh Green's function($G^K_{4\times4}$), namely $\left[ D_{4\times4}^{R(A)}\right] \left( \omega \right) =\left[ G_{4\times 4}^{R(A)}\right] ^{-1}\left(\omega \right) ,G_{4\times 4}^{K}\left(\omega
\right) =-G_{4\times 4}^{R}\left(\omega \right)$ $ D_{4\times 4}^{K}\left(
\omega \right) G_{4\times 4}^{A}\left( \omega \right) .$ The retarded self-energy is a $4\times 4$ matrix

	\begin{equation}
	\sum\nolimits_{4\times 4}^{R}\left( \omega \right) =\left[ 
	\begin{array}{cccc}
	\sum\nolimits_{\left( 1,1\right) }^{R}\left( \omega \right) & 0 & 0 & 0 \\ 
	0 & \sum\nolimits_{\left( 2,2\right) }^{R}\left( \omega \right) & 0 & 0 \\ 
	0 & 0 & \sum\nolimits_{\left( 3,3\right) }^{R}\left( \omega \right) & 0 \\ 
	0 & 0 & 0 & \sum\nolimits_{\left( 4,4\right) }^{R}\left( \omega \right)%
	\end{array}%
	\right],
	\end{equation}%
where

\begin{equation}
\qquad\qquad\sum\nolimits_{\left( 1,1\right) }^{R}\left( \omega \right)
=2iG_{1}^{2}\int_{-\infty }^{+\infty }\frac{d\omega ^{^{\prime }}}{2\pi }%
G_{\left( 3,3\right) }^{K}\left( \omega ^{^{\prime }}\right) G_{\left(
	3,3\right) }^{R}\left( \omega -\omega ^{^{\prime }}\right), \label{eq:retarded-g1}
\end{equation}
\begin{equation}
\qquad\qquad\sum\nolimits_{\left( 2,2\right) }^{R}\left( \omega \right),
=2iG_{2}^{2}\int_{-\infty }^{+\infty }\frac{d\omega ^{^{\prime }}}{2\pi }%
G_{\left( 4,4\right) }^{K}\left( \omega ^{^{\prime }}\right) G_{\left(
	4,4\right) }^{R}\left( \omega -\omega^{^{\prime}}\right)\label{eq:retarded-g2},
\end{equation}
\begin{eqnarray}
\sum\nolimits_{\left( 3,3\right) }^{R}\left( \omega \right)=2iG_{1}^{2}\int_{-\infty }^{+\infty }\frac{d\omega ^{^{\prime }}}{2\pi }\left[ G_{\left( 3,3\right) }^{A}\left( \omega ^{^{\prime }}-\omega \right)\right.&&\left.
G_{\left( 1,1\right) }^{K}\left( \omega ^{^{\prime}}\right)\right.\nonumber\\
&&+\left.G_{\left(3,3\right)}^{K}\left(\omega ^{^{\prime }}\right) G_{\left( 1,1\right)}^{R}\left( \omega ^{^{\prime }}+\omega \right) \right]\label{eq:retarded-g3},
\end{eqnarray}
\begin{eqnarray}
\sum\nolimits_{\left( 4,4\right) }^{R}\left( \omega \right)
=2iG_{2}^{2}\int_{-\infty }^{+\infty }\frac{d\omega ^{^{\prime }}}{2\pi }%
\left[ G_{\left( 4,4\right) }^{A}\left( \omega ^{^\prime }-\omega \right)\right.&&\left.
G_{\left( 2,2\right) }^{K}\left( \omega ^{^{\prime }}\right)\right.\nonumber\\
&&\left.+G_{\left(4,4\right) }^{K}\left( \omega ^{^{\prime }}\right) G_{\left( 2,2\right)
}^{R}\left( \omega ^{^{\prime }}+\omega \right) \right]\label{eq:retarded-g4}.
\end{eqnarray}
In Eq.(\ref{eq:retarded-g1})-Eq.(\ref{eq:retarded-g4}) the entities $G^{A/K/R}_{(j,j)}$ ($j=1,2,3,4$. It represents the element in row $j$ and column $j$ of corresponding Green function matrix form.) can be obtained from matrices in Eq.(\ref{mat:keldysh-g1}) and Eq.(\ref{mat:keldysh-g2}). Note, because the analytical expressions of Eqs.\ref{eq:retarded-g1}, \ref{eq:retarded-g2}, \ref{eq:retarded-g3} and \ref{eq:retarded-g4} are very complicated, they are not listed here. Therefore, we can calculate the full retarded Green’s function $\Im _{4\times 4}^{R}\left( \omega \right) $ and its inverse via the Dyson equation
\begin{equation}
\qquad\qquad\qquad\qquad\Im _{4\times 4}^{R}\left( \omega \right) =\left[ D_{4\times 4}^{R}\left(
\omega \right) -\sum\nolimits_{4\times 4}^{R}\left( \omega \right) \right]
^{-1}\label{eq:retarded-g-function}.
\end{equation}
\section{Optical Response of in the Active-passive Cavity System}\label{sec:sec2}
In this section, we will examine the optical response of the active-passive cavity optomechanical with various parameter of the system, such as the photon-tunneling coupling $J$, effective gain rate of active cavity $\kappa_{2}$, ambient temperature $T$, and so on. Here, we mainly study the following three cases in terms of the considerations related to mechanical membrane:\\
$1)$ there is only one mechanical membrane in the system;\\
$2)$ there are two identical mechanical membranes in the system;\\
$3)$ there are two distinct mechanical membranes in the system.\\ 
Generally, the output field can be obtained by using the standard input-output relation $\varepsilon_{out}+\varepsilon_p e^{-i\delta t}+\varepsilon_{l}=\sqrt{k_1^e}\langle\hat{a}_1\rangle$\cite{57}. Then the total output field at the frequency $\omega_p$ is given by $\varepsilon _{T}=\varepsilon _{out}(\delta )e^{-i\delta t}/\varepsilon
_{p}(\delta )e^{-i\delta t}+1$, here $\varepsilon _{out}(\delta )$ is the optical response at the probe field. In the present paper, we use the method in Refs.\cite{62,63}, the total output field $\varepsilon _{T}$ can be given by the retarded Green’s function. According to Eq.(\ref{eq:retarded-g-function}) we get
\begin{eqnarray}
\varepsilon _{T}&=&i\kappa _{1}^{e}\Im _{\left( 1,1\right) }^{R}\left(
\delta \right)  \nonumber\\
&=&\frac{i\kappa _{1}^{e}\left( \delta -\Delta _{a}-\frac{\kappa _{2}}{2}i-\frac{2G_{2}^{2}\left( n_{2th}+1\right) }{\delta -2\omega _{b2}+i\gamma _{2}}\right) }{\left( \delta -\Delta _{a}-\frac{\kappa _{2}}{2}i-\frac{2G_{2}^{2}\left( n_{2th}+1\right) }{\delta -2\omega _{b2}+i\gamma _{2}}\right) \left( \delta -\Delta _{a}+\frac{\kappa _{1}}{2}i-\frac{2G_{1}^{2}\left( n_{1th}+1\right) }{\delta -2\omega _{b1}+i\gamma _{1}}\right)-J^{2}} \nonumber\\
&=&\mu _{p}+i\nu _{p}.\label{eq:output-field}
\end{eqnarray}
where $\mu_{p}$ and $\nu_{p}$ represent the behaviors of absorption and dispersion coefficient of the active-passive cavity optomechanical system, respectively. Along with that, we can calculate transmission rate using Eq.(\ref{eq:output-field}) as follows
\begin{equation}
\qquad\qquad\qquad\qquad\qquad\left\vert t_{p}\right\vert ^{2}=\left\vert 1-\varepsilon _{T}\right\vert
^{2}.
\end{equation}
Unless stated otherwise, the parameter values chosen here are mainly as follows\cite{33,34,35,36,37,37.1}.
The wavelength of pumping field $\lambda_1$ = $\lambda_2=532$nm, the decay rate of the
passive and active cavity $\kappa_{1}=\kappa_{2}^l=4\pi\times10^4$Hz, the frequency of the membrane
$\omega_{m1}=\omega_{m2}=2\pi\times10^5$Hz and the corresponding effective mechanical frequency
are $\omega_{b1}=\omega_{b2}=5.06\kappa_{1}$, the decay rate of membrane $\gamma_{1}=\gamma_{2}=40$Hz, the
ambient temperature $T=90$K, the effective optomechanical coupling strength
$G_1=G_2=2.81\pi\times10^{-1}$Hz, photon-tunneling coupling $J=0.5\kappa_{1}$, the cavity field detuning $\Delta_{a}=\omega_{m1}+\omega_{m2}$ and normalized optical field detuning $\delta ^{^{\prime }}=\delta
/\omega _{m1}$. 

\subsection{Only One Mechanical Membrane and Two Identical Membrane in the System}
At first, we consider the simplest case, that is the optical response of the system to the weak probe field when there is only one mechanical membrane and two identical mechanical membranes in the active-passive cavity optomechanical system, and compare afterwards analyze them. \\
\begin{figure}[t!]
	\centering\includegraphics[width=0.8\linewidth]{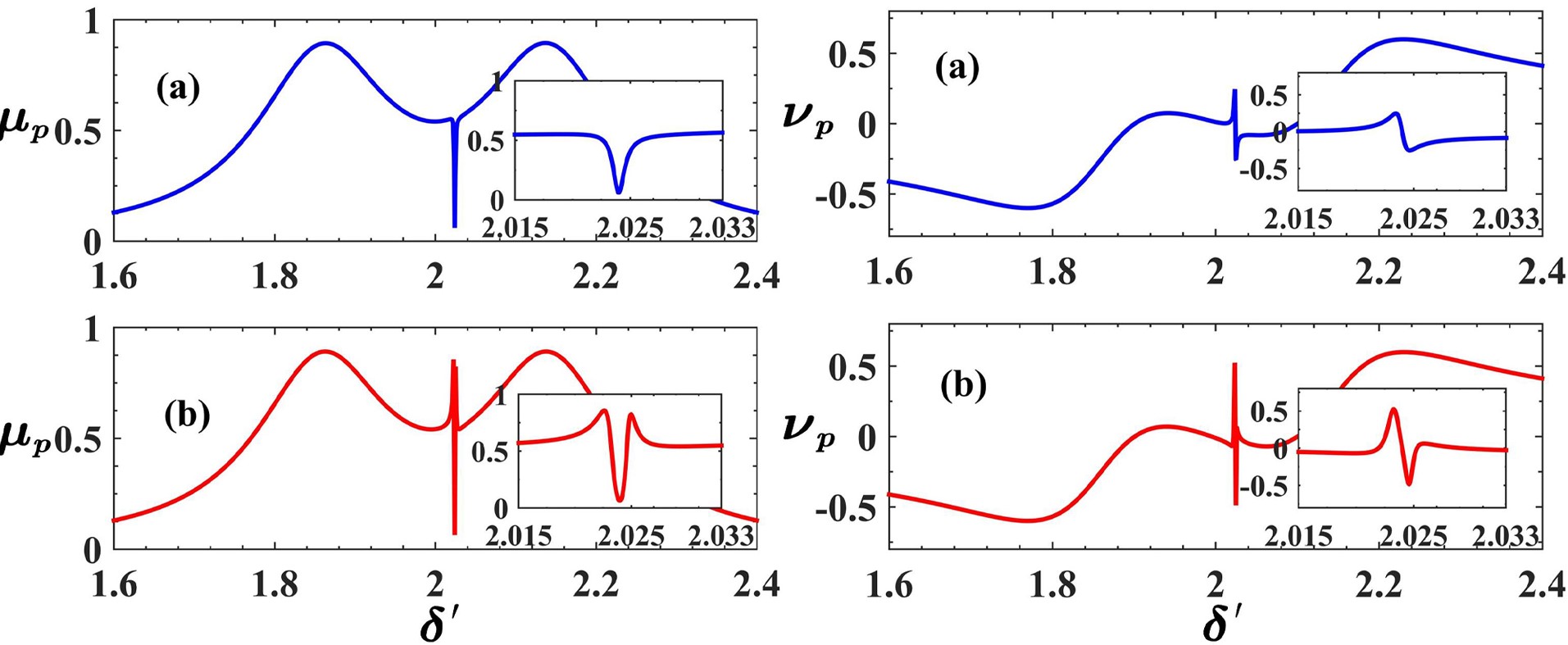}
	\caption{\small The absorption curves $\mu_{p}$ and dispersion curves $\nu_{p}$ as functions of the normalized optical detuning  $\delta^\prime$. The blue curves mean that there is only one mechanical membrane ($g_2=0$) in the active-passive cavity system, and the red curves mean there are two identical mechanical membranes in the system. The photon-tunneling strength between two cavities is  $J=0.7\kappa_{1}$, and $\Delta_{a}=\omega_{m1}+\omega_{m2}=2\omega_{m1}$.}
	\label{fig:absorption-dispersion-curves}
\end{figure}

\begin{figure}[t!]
	\centering
	\includegraphics[width=0.5\linewidth]{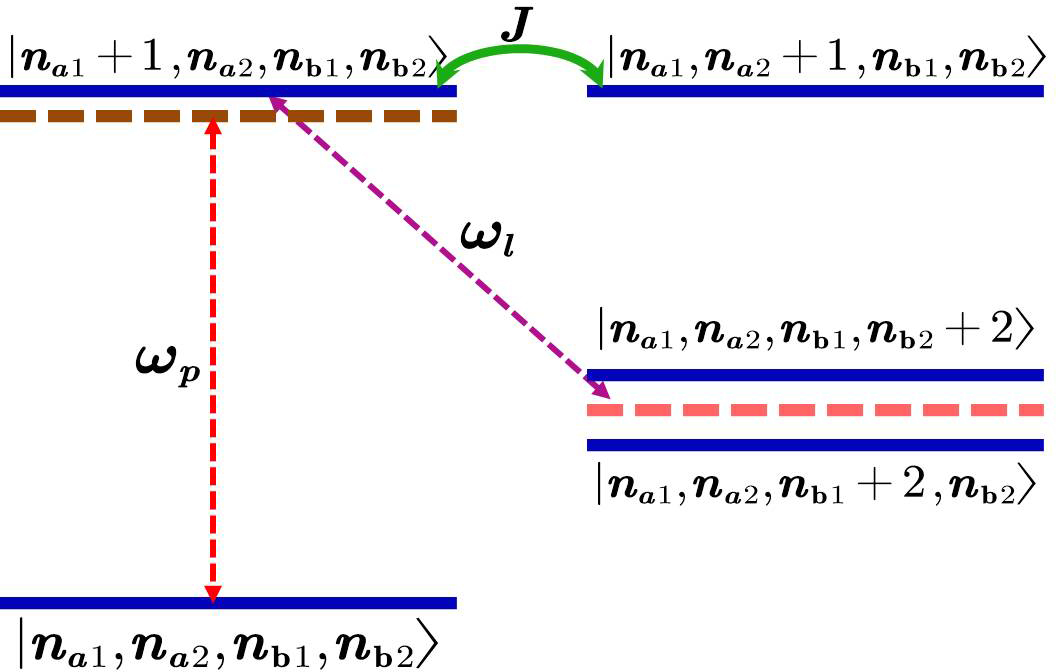}
	\caption{\small Schematic of the energy-level diagram in the active-passive cavity optomechanical system, where $|n_{a1}\rangle$, $| n_{a2}\rangle$, $|n_{b1}\rangle$, $| n_{b2}\rangle$ denote the photon states of the passive cavity mode, active cavity mode, the mechanical membrane of passive cavity (membrane 1) and the mechanical membrane of active cavity (membrane 2), respectively.}
	\label{fig:energy-level-active-passive}
\end{figure}

In Fig.\ref{fig:absorption-dispersion-curves}, we plot the absorption curves $\mu_{p}$ and dispersion curves $\nu_{p}$ as function of the normalized optical detuning $\delta^\prime$ when there is only one mechanical membrane (blue curves) and two identical mechanical membranes (red curves) in the  active-passive cavity system. According to the absorption curves, we can find that when there is only one mechanical membrane in the system, a narrow and deep transparent window appears inside the wide and shallow transparent window. The position of wide and shallow transparent window appears at $\delta^\prime=2$, and the narrow and deep transparent window is at $\delta^\prime=2.024$. This can be understood as when $\delta= \delta^\prime\omega_{m1}=2\omega_{m1}$ and $\Delta_{a}=2\omega_{m1}$, the probe field is originally resonantly absorbed by the passive cavity. However, the normal-mode splitting effect caused by the photon-tunneling coupling between the passive cavity and the active one destroys this resonance absorption. In addition, our system is worked in the resolved-sideband regime, the Stokes scattering is strongly suppressed while the anti-Stokes field (two-phonon process) with frequency $\omega_{l}+2\omega_{b1}$ is promoted. When anti-Stokes field is degenerate with the probe field, these two fields destructively interfere, which lead to OMIT phenomenon at $\delta^\prime=2\omega_{b1}/\omega_{m1}=2.024$. The above physical process can be clearly reflected through the energy-level diagram of Fig.\ref{fig:energy-level-active-passive}.

When there are two identical mechanical membranes in the passive-active cavity optomechanical system, one can find that the position of the narrow and deep transparent window does not change, but two sharp absorption peaks appear near it, which indicates that the transparent window becomes clearer (by comparing the subgraphs of the absorption curve in the two cases). The physical mechanism behind this is the probe field destructively interferes simultaneously with the anti-Stokes field with frequency $\omega_{l}+2\omega_{b1}$ in passive cavity and the anti-Stokes field with frequency  $\omega_{l}+2\omega_{b2}$ ($\omega_{b1}=\omega_{b2}$) (through photon hopping into the passive cavity) in the active cavity, hence the OMIT effect becomes more significant.

From the dispersion curves, we can find that it undergoes a sharp dispersion change at $\delta^\prime=2.024$ when the probe field passing through the active-passive cavity optomechanical system, which plays a significant role in the investigation of slow and fast lights propagation. When two identical mechanical membranes exist at the same time in the system, the dispersion change is more obvious. \\
\begin{figure}[t!]
	\centering
	\includegraphics[width=0.7\linewidth]{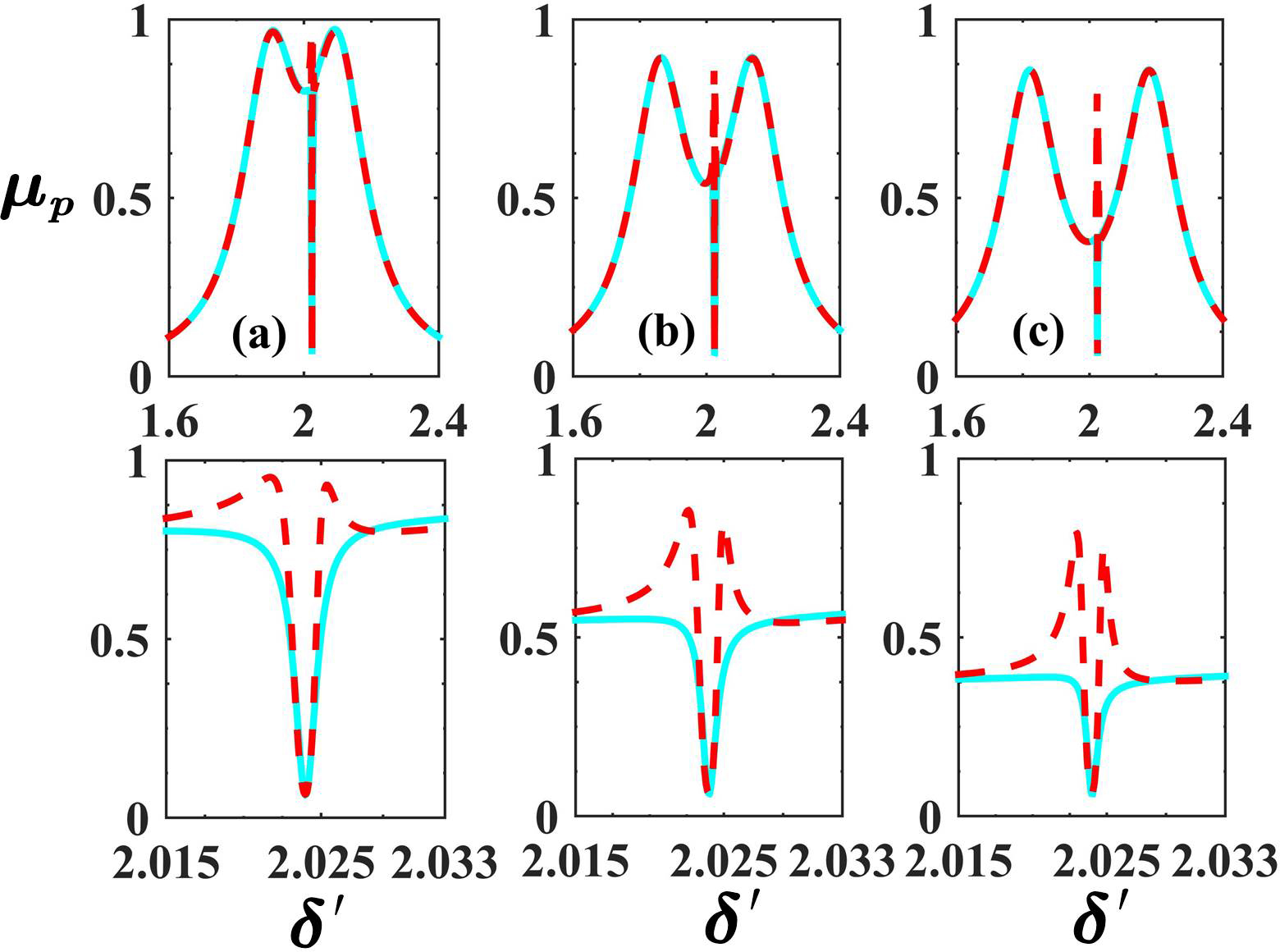}
	\caption{\small The absorption $\mu_{p}$ as functions of the normalized optical detuning $\delta^\prime$ when photon-tunneling strength between two cavities is different, in which the sky blue curves mean only one mechanical membrane, the red curves mean two identical mechanical membrane are present. Figs.4 (a), 4(b) and 4(c) represent  $J=0.5\kappa_{1}$, $0.7\kappa_{1}$ and, $0.9\kappa_{1}$, respectively. The second row represent corresponding subfigures.}
	\label{fig:absorption-with-normalize}
\end{figure}

As shown in Fig.\ref{fig:absorption-with-normalize}, the absorption  $\mu_{p}$ is plotted as a function of normalized optical detuning $\delta^\prime$ for the different photon-tunneling strength $J$. From Figs.\ref{fig:absorption-with-normalize}(a), \ref{fig:absorption-with-normalize}(b) and \ref{fig:absorption-with-normalize}(c), we can easily find that as $J$ increases, the previously wide and shallow transparent window becomes wider and deeper. This is because the increase of photon-tunneling $J$ enhances the normal-mode splitting effect between the two cavity modes, which lead to significant enhancement of the transparency effect. However, according to the subfigures (look at the second row), we can clearly see that with the increase of photon-tunneling, these transparent windows caused by the two-phonon process becomes narrower and shallower, i.e., the two-phonon process is suppressed by photon-tunneling. In particular, the  Fig.\ref{fig:absorption-with-normalize} shows that the red curves (with two identical mechanical membrane) are relatively less affected by photon-tunneling than the sky blue curves (with one mechanical membrane). This originates from the fact that, comparing with the case of only one membrane in the system, when there are two identical mechanical membrane, part of the anti-Stokes field with frequency$\omega_{l} + 2\omega_{b2}$ ($\omega_{b1}=\omega_{b2}$) in the active cavity enters the passive cavity through photon-tunneling and destructively interferes with the probe field. Therefore, when the photon-tunneling is larger, the difference between the red curves and the sky blue curves becomes more significant. It should be noted that this explanation is not inconsistent with the conclusion that OMIT effect of the red curve is reduced, because more the anti-Stokes field may from the passive cavity leaks to the active cavity. \\

\begin{figure}[t!]
	\centering\includegraphics[width=0.6\linewidth]{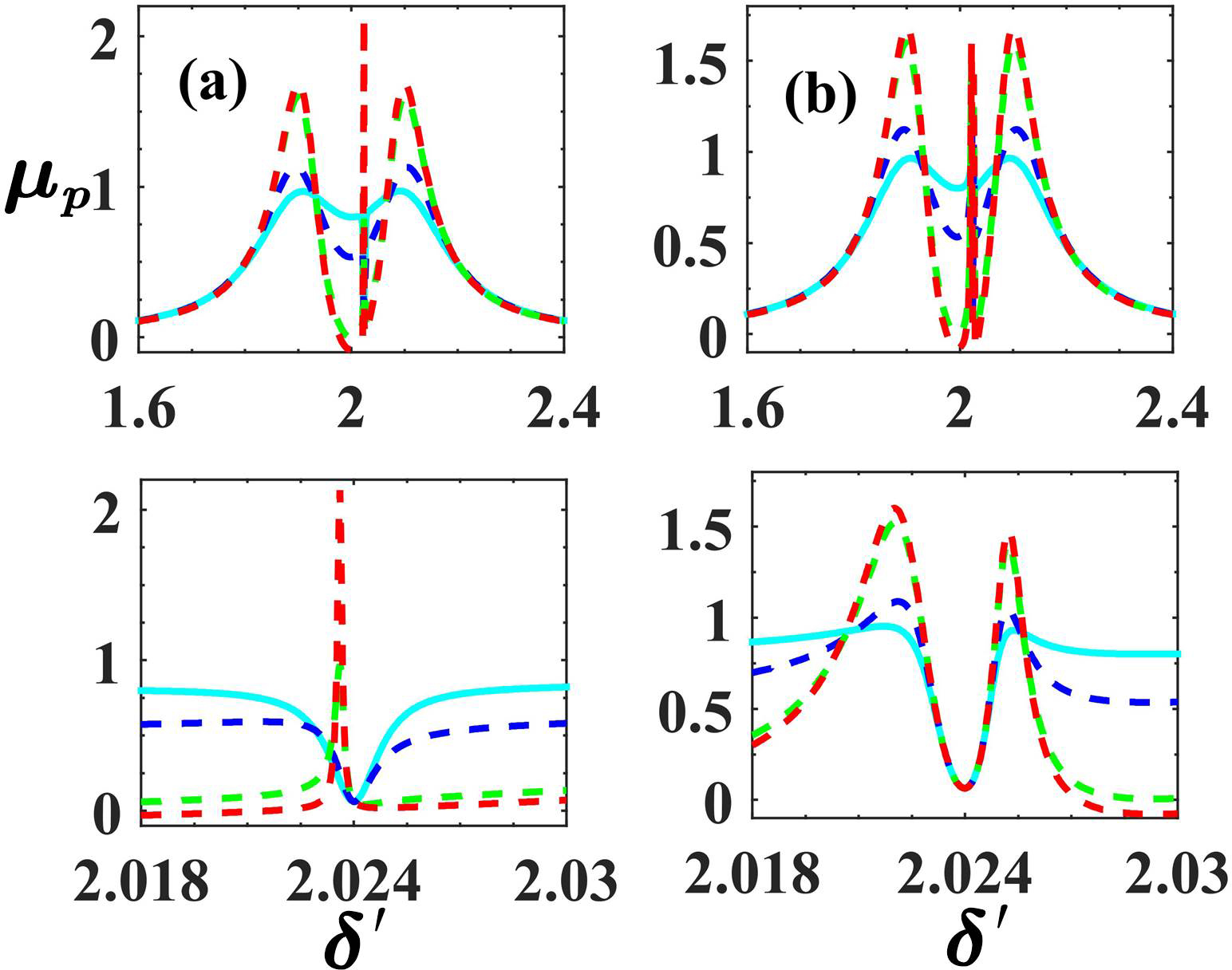}
	\caption{\small The absorption  $\mu_{p}$ as functions of the normalized optical detuning $\delta^\prime$ for different the effective pump rate of active cavity. In Fig.5(a) there is only one mechanical membrane (\textit{i.e.} $g_2=0$) in the system, and in Fig.5(b) there are two identical mechanical membranes. The second row represents the corresponding subfigures. Among them, sky blue, blue, green and red curves represent $\kappa_{2}=-\kappa_{1}$, $-0.5\kappa_{1}$, $0$, $0.05\kappa_{1}$, respectively. }
	\label{fig:absorption-normalize-optical-detuning}
\end{figure}
In Fig.\ref{fig:absorption-normalize-optical-detuning}, the absorption $\mu_{p}$ of the output field depending on normalized optical detuning $\delta^\prime$ is showed for the effective pump rate of active cavity  $\kappa_{2}$. Figs. \ref{fig:absorption-normalize-optical-detuning}(a) and \ref{fig:absorption-normalize-optical-detuning}(b) indicate that as $\kappa_{2}$ increases, the depth of transparent window caused by photon-tunneling gradually becomes deeper and then remains unchanged, but the width never changes significantly (this is different from the case of increasing $J$). This is due to the increase of  $\kappa_{2}$ only enhances the effectively photon-tunneling between two cavities  but does not improve the standard normal-mode splitting (the supermode frequency formed by two cavity modes only depends on $J$), so the depth of the transparent window increases, but the width does not noticeably change. 

Surprisingly, according to the subfigure of Fig.\ref{fig:absorption-normalize-optical-detuning}(a), we can find that when there is only one mechanical membrane in the system, as $\kappa_{2}$ increases, the transparent window caused by the two-phonon process gradually becomes shallower and then converted into an absorption peak, and the absorption coefficient can be up to 2 or more. The physical mechanism behind this is that the OMIT caused by two-phonon process is destroyed when the effective pump rate of active cavity $\kappa_{2}$ increased, and the gain effect will play a dominant role. However, from the subfigure of Fig. \ref{fig:absorption-normalize-optical-detuning}(b), we can see that when there are two mechanical membranes in the system, with the increase of $\kappa_{2}$, the OMIT effect caused by the two-phonon process is enhanced. This is due to the fact that the increase of $\kappa_{2}$ promotes the photon-tunneling between two cavities, thereby enhancing the destructive interference between the probe field and the anti-Stokes field in the active cavity, which result in an enhanced OMIT effect.\\

\begin{figure}[t!]
	\centering\includegraphics[width=0.6\linewidth]{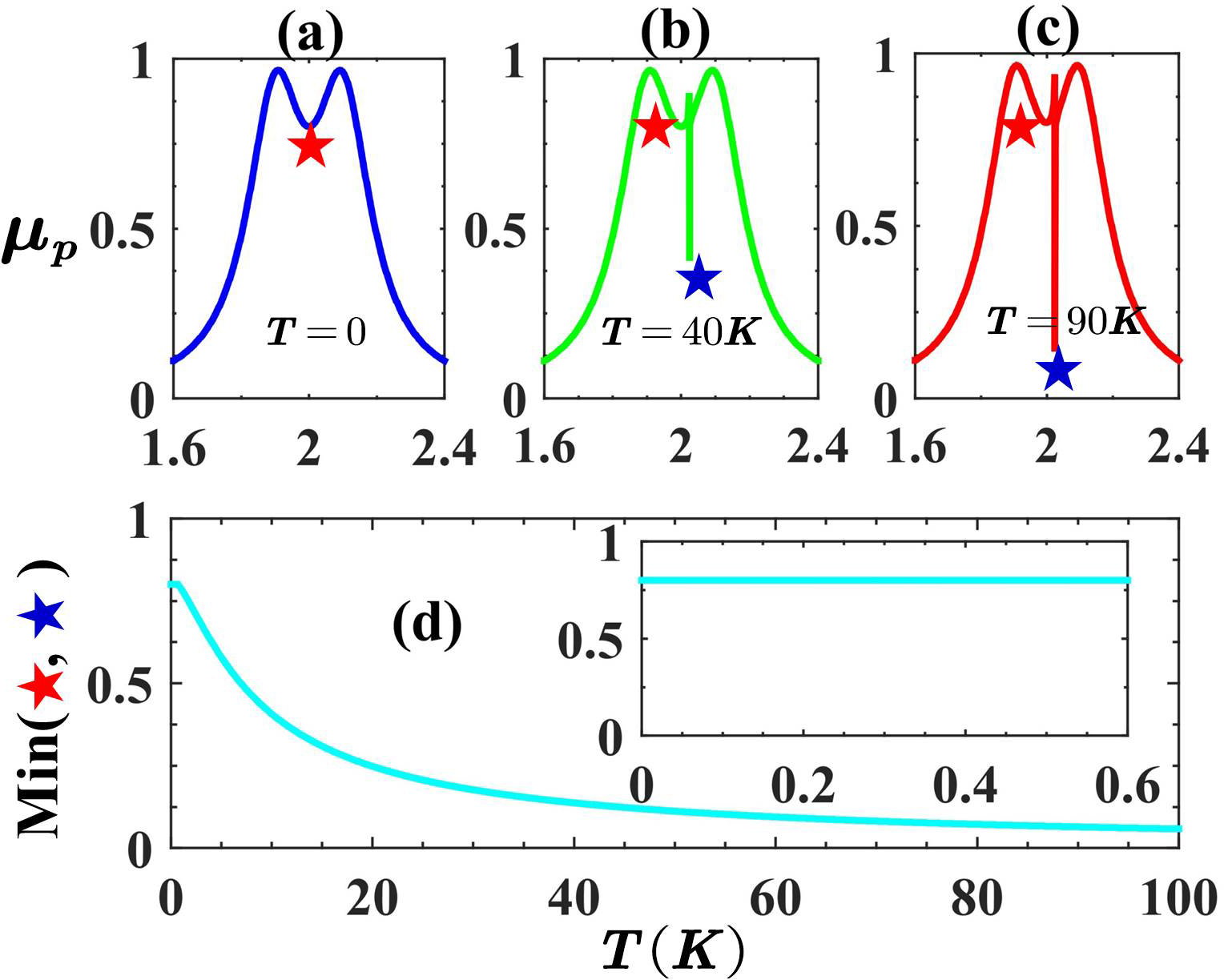}
	\caption{\small In Figs.6 (a), 6(b) and 6(c), the absorption  $\mu_{p}$ as a function of normalized optical detuning $\delta^\prime$ for different ambient temperatures has been observed when there are two identical mechanical membranes in the active-passive cavity system. Fig.6(d) shows that the change of the minimum absorption value of the system with environment temperature $T$.  }
	\label{fig:absorption-coeff-with-normalixed-detuning}
\end{figure}
We presented in Figs.\ref{fig:absorption-coeff-with-normalixed-detuning}(a), \ref{fig:absorption-coeff-with-normalixed-detuning}(b) and \ref{fig:absorption-coeff-with-normalixed-detuning}(c) are the absorption coefficient depending on the normalized optical detuning $\delta^\prime$ under the different environment temperatures when there are two identical mechanical membranes in the passive-active cavity system. In Fig.\ref{fig:absorption-coeff-with-normalixed-detuning}(d), we observe the change of the minimum absorption value of transparency window (location marked by a five-pointed star) with environment temperature. From Figs.\ref{fig:absorption-coeff-with-normalixed-detuning}(a), \ref{fig:absorption-coeff-with-normalixed-detuning}(b) and \ref{fig:absorption-coeff-with-normalixed-detuning}(c), we can conclude that as the ambient temperature $T$ increases, the transparent window caused by photon-tunneling does not change significantly, but the OMIT effect caused by the two-phonon process becomes more significant, i.e., the transparent window becomes deeper. This is because the ambient temperature is introduced into the system through the phonon channel, which cause the thermal vibration of mechanical membranes and promoting the two-phonon process, hence the OMIT effect becomes obvious. 

According to Fig \ref{fig:absorption-coeff-with-normalixed-detuning}(d), we find that when the temperature increases from 0 to 0.6K, the minimum absorption value of transparent window basically does not change. However, after we increase the temperature more than 0.6K, the minimum absorption value of transparent window decreases rapidly, then tends to stabilize close to 0. This originate from the fact when the temperature is very low, the minimum absorption is caused by the photon-tunneling, and when the temperature is high, the minimum absorption is caused by the two-phonon process, 0.6K is the corresponding critical temperature.

\subsection{Two Different Mechanical Membranes in the System}
In this subsection, we will study a more general case, namely response of the active-passive cavity optomechanical system with two different membranes to a weak probe field. Here, we mainly observe the change of transmission coefficient of the system with various physical parameters.\\

\begin{figure}[t!]
	\centering\includegraphics[width=0.6\linewidth]{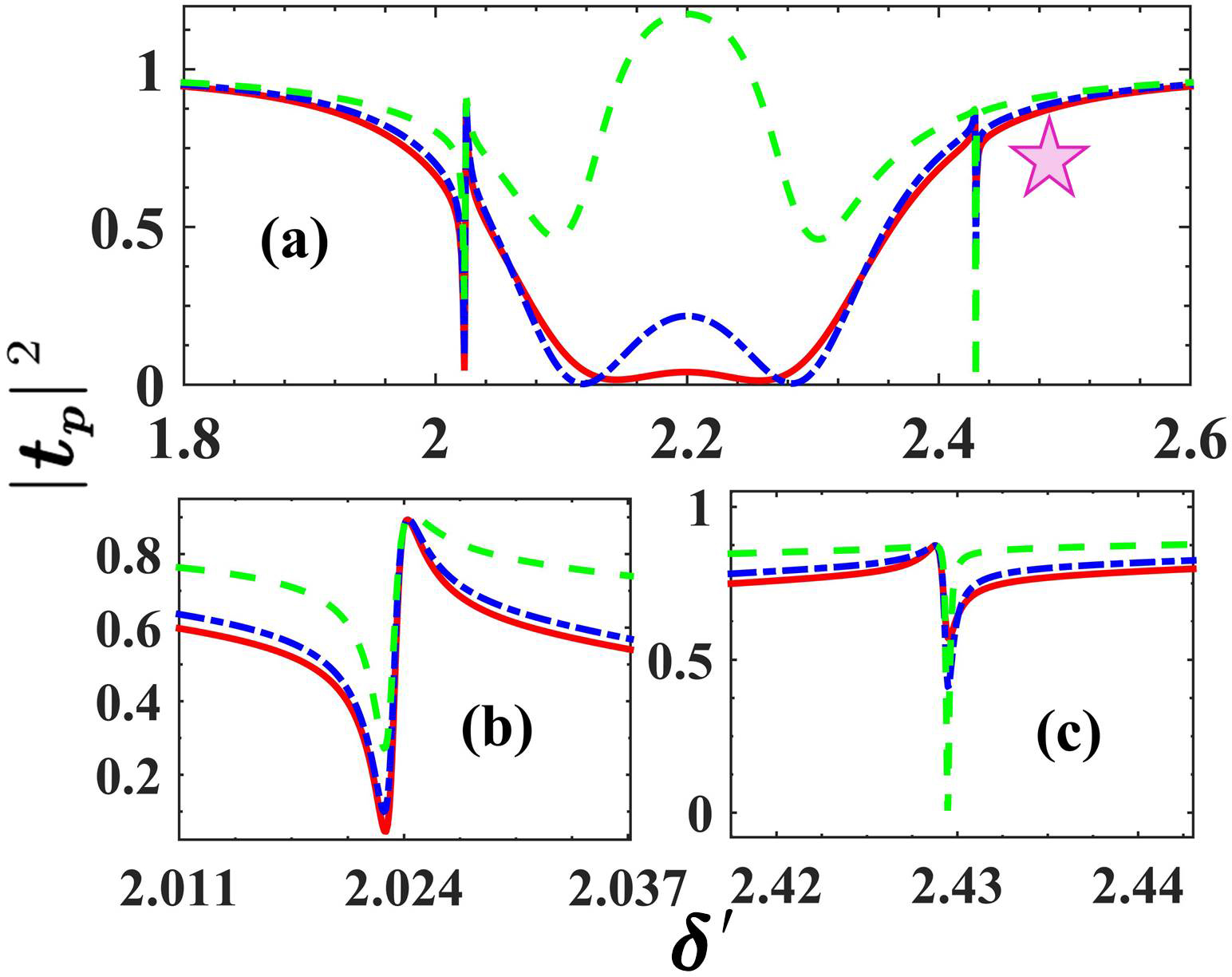}
	\caption{\small The transmission coefficient $|t_p|^2$ varies with normalized optical detuning $\delta^\prime$ for different the effective pump rate of active cavity, here (b) and (c) are corresponding subfigures. In which red, blue and green curves indicate that $\kappa_{2}=-\kappa_{1}$, $-0.5\kappa_{1}$, $0.05\kappa_{1}$, respectively. The frequency of the mechanical membrane 2 is $\omega_{m2}=1.2\omega_{m1}$, and corresponding effective mechanical frequency is $\omega_{b2}=1.2\omega_{b1}$. }\label{fig:transmission-coefficient-with-normalized-detuning}
\end{figure}

Fig.\ref{fig:transmission-coefficient-with-normalized-detuning} plots the transmission coefficient $|t_p|^2$ as a function of the normalized optical detuning $\delta^\prime$ for different the effective pump rate of active cavity $\kappa_{2}$. We can easily see that there is one more transparent window (see the five-pointed star) in the transmission spectrum compared to the two cases discussed earlier, that is, three transparent windows. Obviously, this originates from different effective mechanical frequencies of the two mechanical membranes. Among them, the wide transparent window in the middle is caused by the normal-mode splitting by photon-tunneling, and its position appears in $\delta^\prime=\Delta_{a}/\omega_{m1}=2.2$. The two narrow transparent windows on the left and right are caused by the two-phonon process, and their position appear in $\delta^\prime=2\omega_{b1}/\omega_{m1}=2.024$ and $\delta^\prime=2\omega_{b2}/\omega_{m1}=2.4288$, respectively. In particular, we find that with the increase of $\kappa_{2}$, the transparent windows on the left and the middle move up, which means that the transmittance of system at these two places increases. However, in this case, strong absorption appears in the right window (the green curve in Fig.7(c)). This can be understood as the increase in $\kappa_{2}$, which promotes the anti-Stokes field with frequency $\omega_{l}+2\omega_{b2}$ in the active cavity enter the passive cavity, and then destructively interfere with the probe field of the same frequency, thus increasing absorption of the weak probe field. \\

\begin{figure}[t!]
	\centering\includegraphics[width=0.6\linewidth]{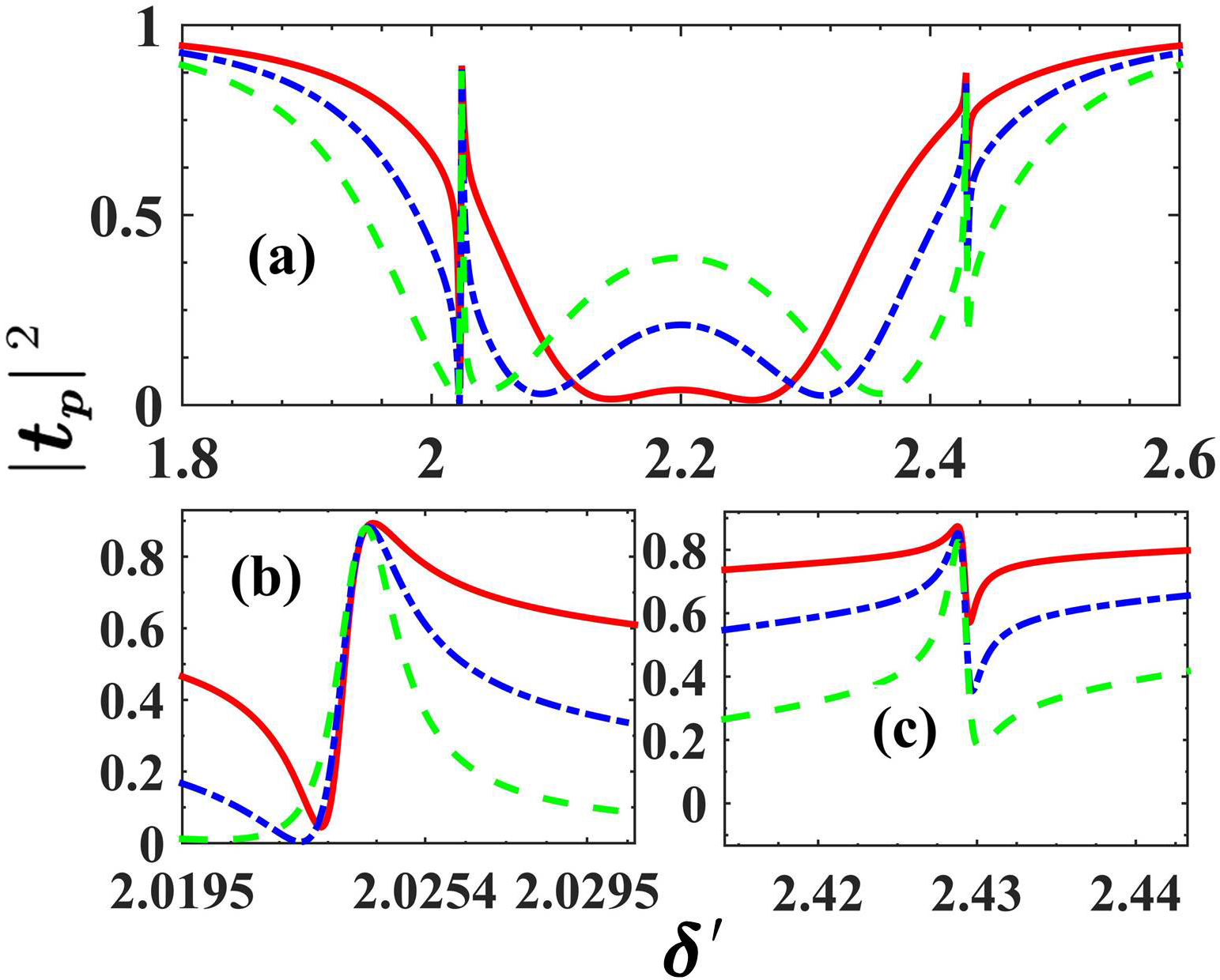}
	\caption{\small The transmission coefficient $|t_p|^2$ as functions of the normalized optical detuning $\delta^\prime$ for different photon-tunneling strength between two cavities $J$. In which red, blue and green curves indicate that $J=0.5\kappa_{1}$, $0.7\kappa_{1}$, $0.9\kappa_{1}$, respectively. The frequency of the mechanical membrane 2 is $\omega_{m2}=1.2\omega_{m1}$, and corresponding effective mechanical frequency is $\omega_{b2}=1.2\omega_{b1}$.}
	\label{fig:transmission-coefficient-with-normalized-detuning-j-vary}
	\end{figure}

In Fig.\ref{fig:transmission-coefficient-with-normalized-detuning-j-vary}, the transmission coefficient $|t_p|^2$ is plotted as function of the normalized optical detuning $\delta^\prime$ for different the photon-tunneling strength $J$, here Figs.\ref{fig:transmission-coefficient-with-normalized-detuning-j-vary}(b) and \ref{fig:transmission-coefficient-with-normalized-detuning-j-vary}(c) are corresponding subfigures. We can find that, as photon-tunneling strength $J$ increases, the transparent window caused by photon-tunneling becomes wider and deeper, which is consistent with the previous conclusion. In addition, we can see that the maximum transmittance of the transparent window caused by the two-phonon process did not change significantly, and the transmittance near the transparent window on the right has decreased significantly. This can be understood as follows. An increment in $J$ causes tunnelling of more anti-Stokes fields with frequency $\omega_{l}+2\omega_{b2}$ of the active cavity into the passive cavity through photon-tunneling, and then it destructively interferes with the probe field of same frequency.\\

\begin{figure}[t!]
	\centering\includegraphics[width=0.6\linewidth]{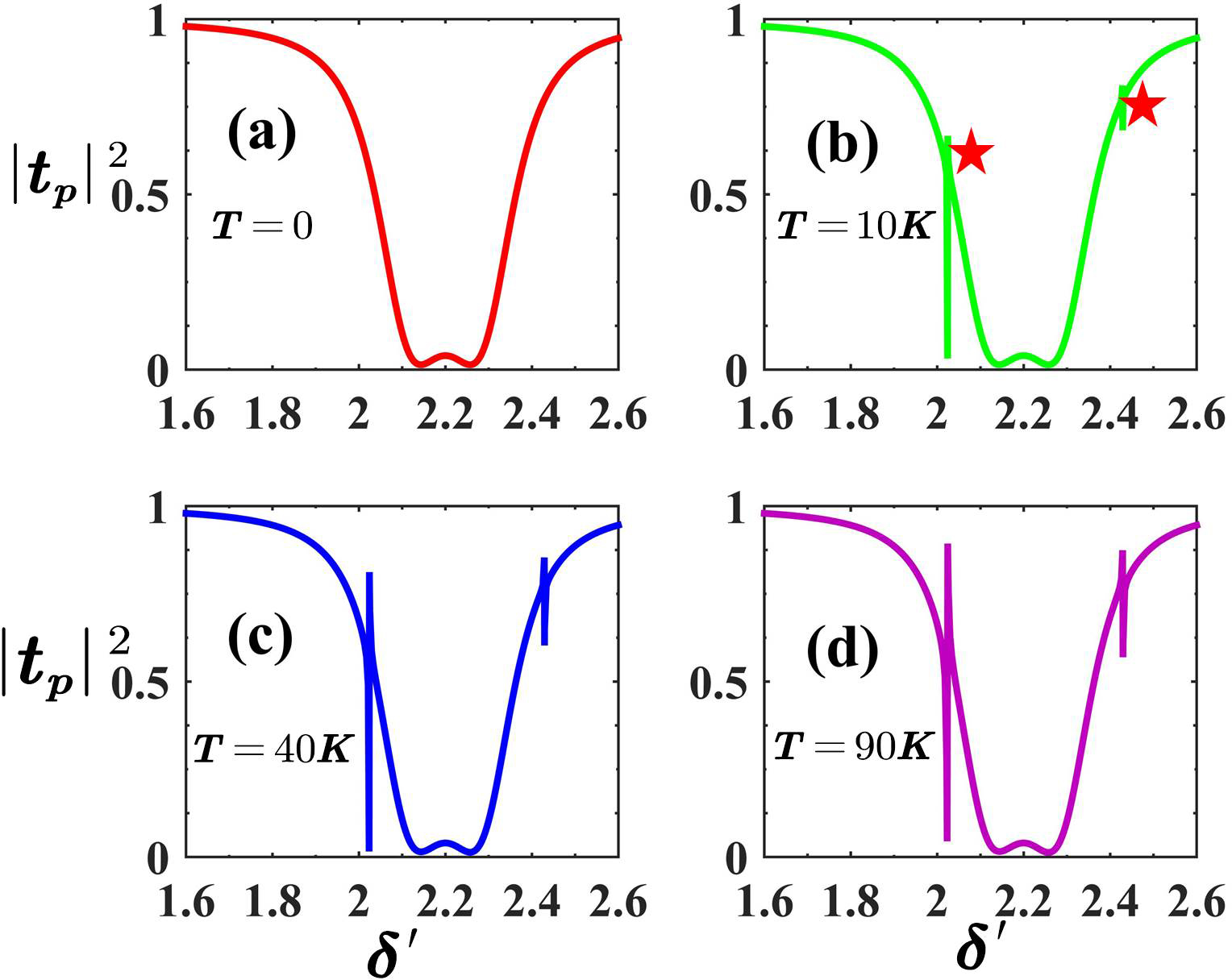}
	\caption{\small The transmission coefficient $|t_p|^2$ as functions of the normalized optical detuning $\delta^\prime$ for different ambient temperatures when there are two different mechanical membranes in the active-passive cavity system. The frequency of the mechanical membrane 2 is $\omega_{m2}=1.2\omega_{m1}$, and corresponding effective mechanical frequency is $\omega_{b2}=1.2\omega_{b1}$.}
	\label{fig:transmission-coeff-with-normalized-optical-detuning-temp-vary}
\end{figure}

As shown in Fig.\ref{fig:transmission-coeff-with-normalized-optical-detuning-temp-vary}, the transmission coefficient $|t_p|^2$ varies with normalized optical detuning $\delta^\prime$ for different ambient temperatures when there are two different mechanical membranes in the active-passive cavity system. One can see that when the ambient temperature is equal to zero, there is only one transparent window in the transmission spectrum caused by photon-tunneling. However, when the temperature is not equal to zero, it will cause thermal vibration of the membrane in the two cavities which induces the other two transparent windows (see the five-pointed star in Fig.\ref{fig:transmission-coeff-with-normalized-optical-detuning-temp-vary}(b)). In addition, we can clearly see that the transparent window on the left is more prominent than the transparent window on the right. This is because only part of the anti-Stokes field in the active cavity enters the passive cavity through photon-tunneling, and destructively interferes with probe field. In other words, if $J$ is equal to 0, the window on the right will never appear. In particular, when the temperature increases, the left and right transparent windows gradually become deeper, but the transparent window in the middle not changes, which is obvious. This once again shows that the ambient temperature plays a key role in the two-phonon process of quadratically coupling optomechanical system.\\

\begin{figure}[t!]
	\centering\includegraphics[width=0.6\linewidth]{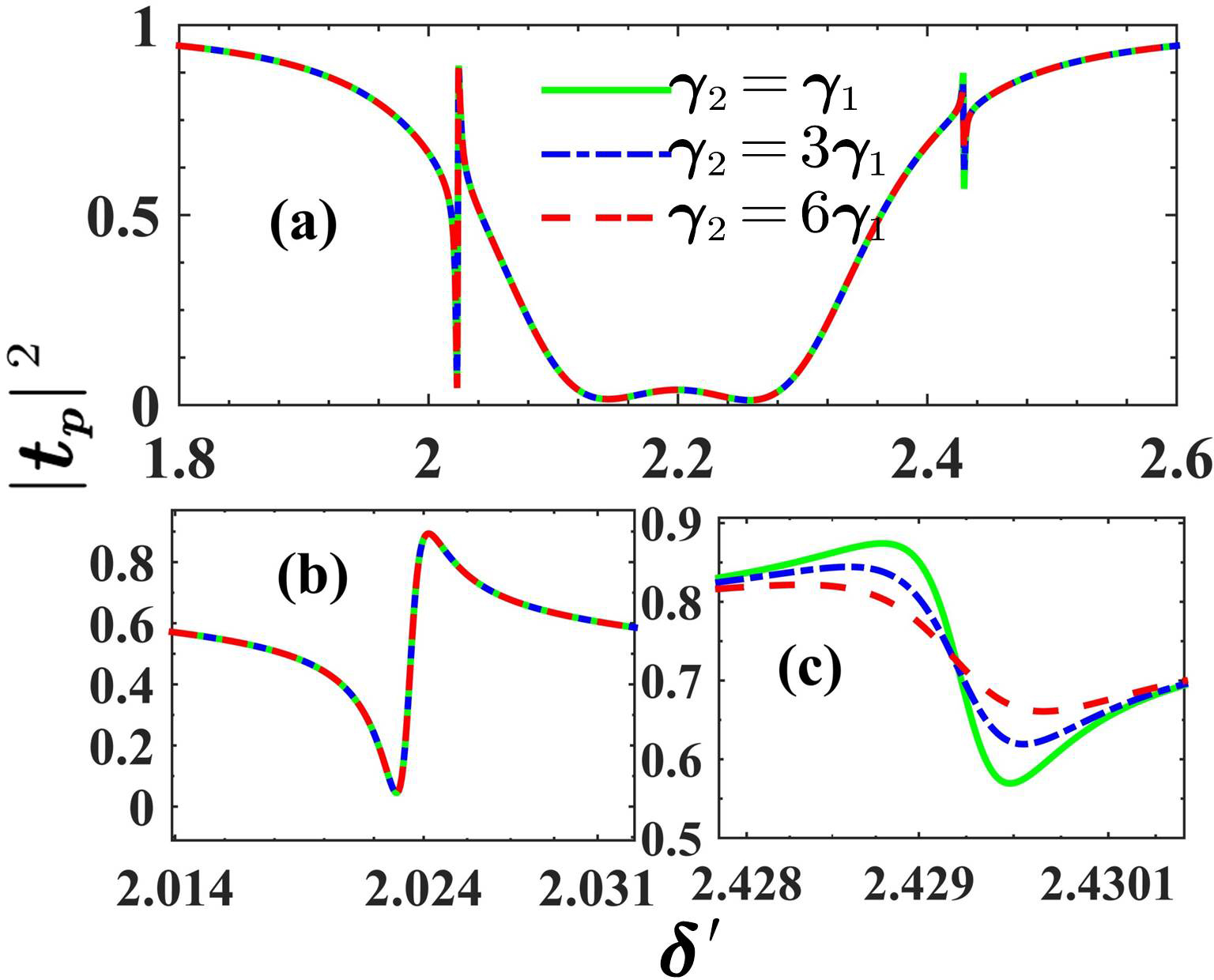}
	\caption{\small The power transmission coefficient $|t_p|^2$ as functions of the normalized optical detuning $\delta^\prime$ for different damping of the mechanical membrane $2$, here ($b$) and ($c$) are the corresponding subfigures. The frequency of the mechanical membrane 2 is $\omega_{m2}=1.2\omega_{m1}$, and corresponding effective mechanical frequency is $\omega_{b2}=1.2\omega_{b1}$}
	\label{fig:transmission-coeff-with-normalized-detuning-vary-dumping}
\end{figure}

Presented in Fig.\ref{fig:transmission-coeff-with-normalized-detuning-vary-dumping} is the transmission coefficient $|t_p|^2$ depending on the normalized optical detuning $\delta^\prime$ under different damping of the mechanical membrane 2. We can find that, with the increase of damping of the mechanical membrane 2, the OMIT effect of the right transparent window is weakened, and the other two transparent windows have no obvious changes. This is because the damping of the mechanical membrane 2 only has a harmful effect on the destructive interference between the probe field and anti-Stokes field with frequency $\omega_{l}+2\omega_{b2}$ and completely damages such an interference effect for the very large $\gamma_{2}$. In addition, we have also considered the influence of $\gamma_{1}$ on the transmission spectrum. We can get a similar conclusion that the OMIT effect of the transparent window on the left is weakened. In order to avoid figure duplication, we will not present it here.
.
\section{Fast and Slow Light in the Active-passive Cavity System}\label{sec:sec3}
In the previous section, we studied the OMIT phenomenon in the active-passive cavity system and found that the dispersion curve appeared a sharp dispersion change near the transparent window, which provided the basis for our next study of fast and slow light identification. Accompanying the OMIT process, the fast or slow light effect also emerge as the optical group delay 
\begin{equation}
\qquad\qquad\qquad\qquad\qquad\qquad\tau _{g}=\frac{\textrm{d}\phi \left( t_{p}\right) }{\textrm{d}\omega _{p}},\label{eq:fast-slow-light-relation}
\end{equation}
where $\phi(t_p)=$Arg$[t_p]$.
\begin{figure}[t!]
	\centering\includegraphics[width=0.6\linewidth]{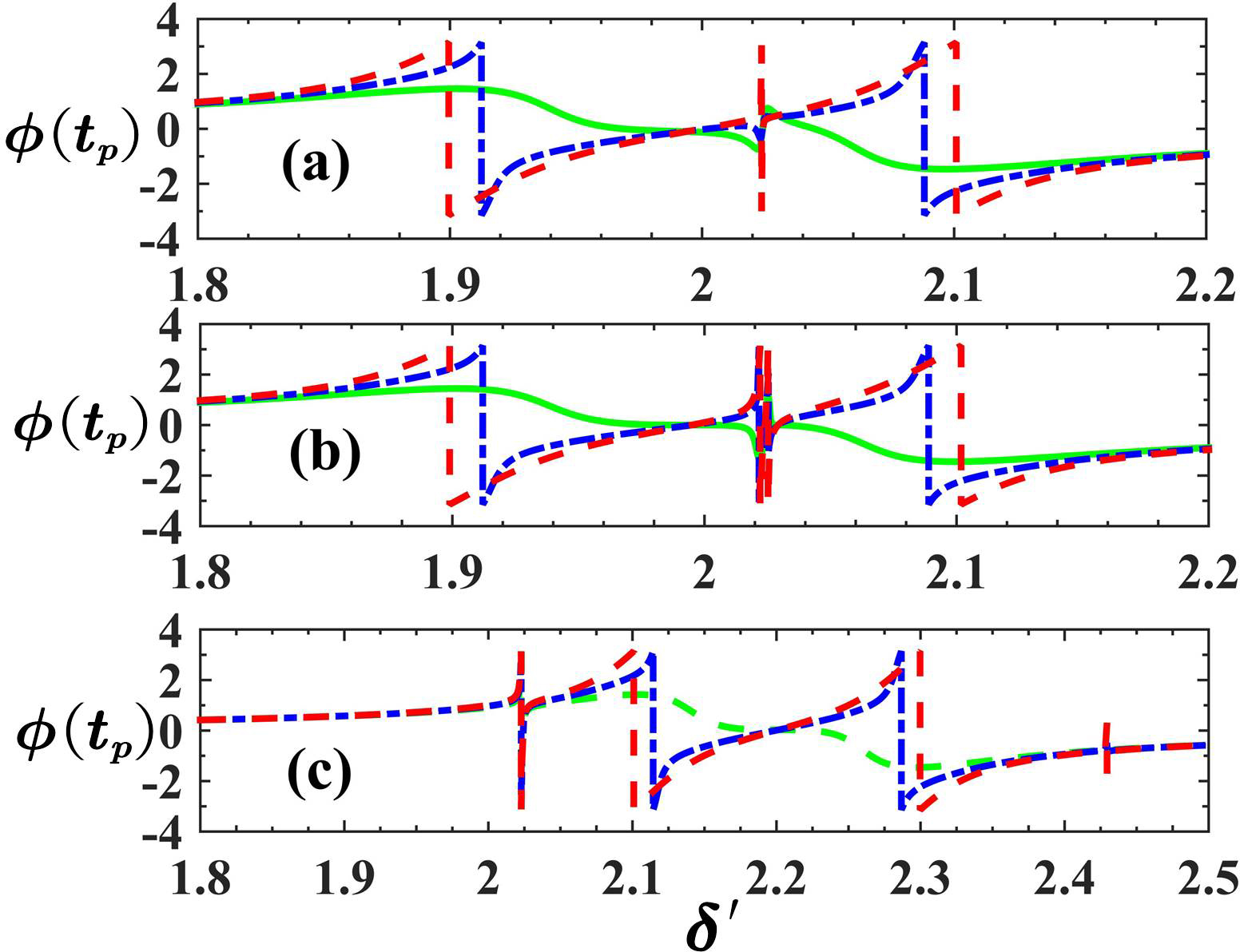}
	\caption{\small The phase dispersion $\phi$($t_p$) as a function of the normalized optical detuning $\delta^\prime$ for different effective pump rate of active cavity. In which ($a$), ($b$) and, ($c$) represent only one mechanical membrane ($g_2=0$), two identical mechanical membranes ($\omega_{m1}=\omega_{m2}$, $\omega_{b1}=\omega_{b2}$) and, two different mechanical membranes ($\omega_{m2}=1.2\omega_{m1}$, $\omega_{b2}=1.2\omega_{b1}$) respectively. The green, blue and red curves indicate that $\kappa_{2}=-\kappa_{1}$, $-0.5\kappa_{1}$ and $0.05\kappa_{1}$, respectively.}
	\label{fig:phase-dispersion-with-normalized-optical-}
\end{figure}
Fig.\ref{fig:phase-dispersion-with-normalized-optical-} shows that the rapid phase dispersion $\phi$($t_p$) as a function of the normalized optical detuning $\delta^\prime$ for different effective pump rate of active cavity $\kappa_{2}$. We can see
that when $\kappa_{2}=-\kappa_{1}$ (see green curves), in Fig.\ref{fig:phase-dispersion-with-normalized-optical-}($a$) and  Fig.\ref{fig:phase-dispersion-with-normalized-optical-}($b$), the phase dispersion $\phi$($t_p$) changes drastically  from negative to positive at $\delta^\prime=2.024$. In Fig.\ref{fig:phase-dispersion-with-normalized-optical-}($c$), $\phi$($t_p$) changes from negative to positive in both $\delta^\prime=2.024$ and $\delta^\prime=2.4288$ at the same time. These positions correspond exactly to the positions of the transparent windows caused by the two-phonon process in three situations discussed earlier. In addition, we can also find that, in Fig.\ref{fig:phase-dispersion-with-normalized-optical-}($a$) and Fig.\ref{fig:phase-dispersion-with-normalized-optical-}($b$), the $\phi$($t_p$) changes sharply around $\delta^\prime=1.9$ and $\delta^\prime=2.1$; and in Fig.\ref{fig:phase-dispersion-with-normalized-optical-}($c$), the $\phi$($t_p$) changed drastically around $\delta^\prime=2.1$ and $\delta^\prime=2.3$. In fact, the difference between the two positions corresponds to the width of the transparent window caused by photon-tunneling under the current $\kappa_{2}$. At the same time, the amplitude of $\phi$($t_p$) change caused by the two-phonon process is also adjusted by $\kappa_{2}$,  but the position (where the dispersion $\phi$($t_p$) changes drastically) remains unchanged. This point corresponds to that $\kappa_{2}$ only changes the depth of the transparent window caused by the two-phonon process, but does not change its position. We know the sharp change of phase dispersion $\phi$($t_p$) near the transparent window which leads to huge change in the refractive index of the medium, thus producing group delay and group advanced phenomenon. Next, we will explore the impact of the effective pump rate of active cavity $\kappa_{2}$ on the fast and slow light in our proposed active-passive cavity optomechanical system. \\

\begin{figure}[t!]
	\centering\includegraphics[width=0.6\linewidth]{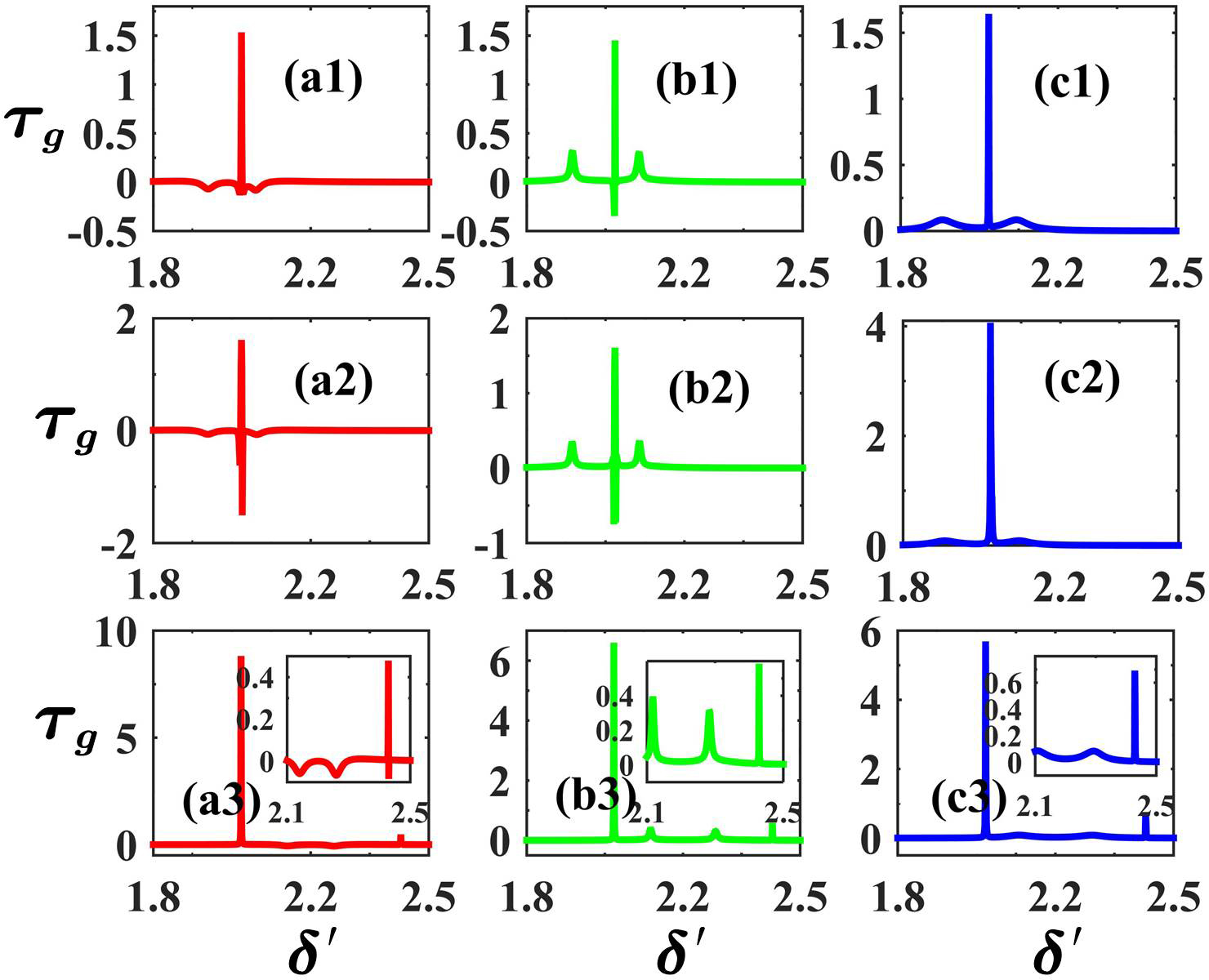}
	\caption{\small The group speed $\tau_{g}$ (in ms) as a function of the normalized optical detuning $\delta^\prime$ for different effective pump rate of active cavity, where the marks $1$, $2$ and $3$ represent only one
		mechanical membrane ($g_2=0$), two identical mechanical membranes ($\omega_{m1}=\omega_{m2}$, $\omega_{b1}=\omega_{b2}$) and two different mechanical membranes ($\omega_{m2}=1.2\omega_{m1}$, $\omega_{b2}=1.2\omega_{b1}$), respectively. The red, green and blue curves indicate that $\kappa_{2}=-\kappa_{1}$, $-0.5\kappa_{1}$ and $0.05\kappa_{1}$, respectively.}
	\label{fig:group-speed-with-normalized-optical-detuning}
\end{figure}
In Fig.\ref{fig:group-speed-with-normalized-optical-detuning}, the group speed $\tau_{g}$ (in ms) is plotted as a function of the normalized optical detuning $\delta^\prime$ for different effective pump rate, $\kappa_{2}$ of active cavity.   Fig.\ref{fig:group-speed-with-normalized-optical-detuning} shows that, when $\kappa_{2}=-\kappa_{1}$ (see red curves), the  active-passive cavity system has fast ($\tau_{g}<0$) and slow ($\tau_{g}>0$) light phenomena in the three situations we studied, and this phenomenon is more prominent near the transparent window. This originates from the fact that the refractive index of the medium is greatly changed near the transparent window, so the fast and slow light effect is obvious. In addition, we find that the maximum group delay in Fig.\ref{fig:group-speed-with-normalized-optical-detuning}($a1$) and Fig.\ref{fig:group-speed-with-normalized-optical-detuning}($a2$) can only reach about $2$ms while in Fig.\ref{fig:group-speed-with-normalized-optical-detuning}($a3$) it can reach $8$ms. When $\kappa_{2}$ increases to $-0.5\kappa_{1}$ (see the green curves), we find that not only the amplitude of the fast and slow light achieved by the system has been adjusted, but also the transition from fast light to slow light can be achieved. This can be reflected by the change of the two short peaks (from negative to positive) in the green curve. This means that the gain effect can strongly modify dispersion of the system. In particular, we see that continue to increase $\kappa_{2}$ (see the blue curves), the fast light effect caused by the two-phonon process is suppressed. This result can be obtained from the change of the peak at $\delta^\prime=2.024$ (corresponds to the highest peak in Fig.\ref{fig:group-speed-with-normalized-optical-detuning}($c1$) and Fig.\ref{fig:group-speed-with-normalized-optical-detuning}($c2$)) and the peak at $\delta^\prime=2.4288$ (corresponds to the rightmost peak in Fig.\ref{fig:group-speed-with-normalized-optical-detuning}($c3$)) with $\kappa_{2}$.\\

\begin{figure}[t!]
	\centering\includegraphics[width=\linewidth]{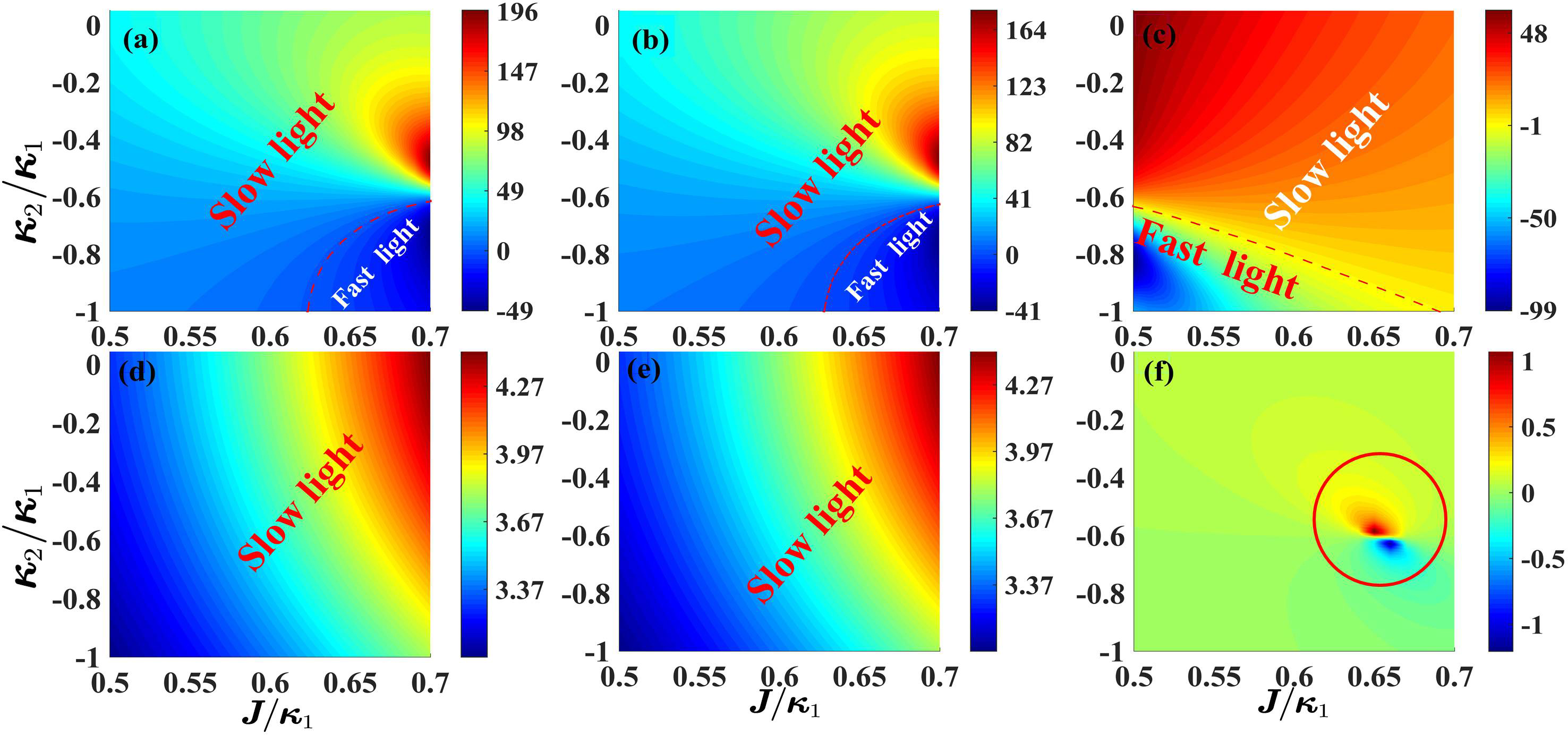}
	\caption{\small The group speed $\tau_{g}$ as a function of the photon-tunneling strength $J$ and effective pump rate of active cavity $\kappa_{2}$. In which Figs.13($a$), 13($b$) and 13($c$) represent only one mechanical membrane ($g_2=0$), two identical mechanical membranes ($\omega_{m1}=\omega_{m2}$, $\omega_{b1}=\omega_{b2}$) and two different mechanical membranes ($\omega_{m2}=1.2\omega_{m1}$, $\omega_{b2}=1.2\omega_{b1}$) respectively for $\delta^\prime=2.136$. The Figs. 13($d$), 13($e$) and 13($f$) has a similar description, just changed $\delta^\prime=2.136$ to $\delta^\prime=2.316$. Here, except that the unit of the Fig.($f$) is in ms, the unit of the other figures is in $\mu$s.}
	\label{fig:group-speed-with-J}
\end{figure}

In Fig.\ref{fig:group-speed-with-J}, density plot of the group speed $\tau_{g}$ versus the photon-tunneling strength $J$ and the effective pump rate of active cavity $\kappa_{2}$. According to Figs.\ref{fig:group-speed-with-J}($a$), 13($b$) and 13($c$), we can find that, when the normalized optical detuning $\delta^\prime=2.136$, Fig.\ref{fig:group-speed-with-J}($a$) and Fig.\ref{fig:group-speed-with-J}($b$) have similar line shapes, that is, the system can achieve fast light in the $-\kappa_{1}<\kappa_{2}<-0.6\kappa_{1}$, $0.63\kappa_{1}<J<0.7\kappa_{1}$, and slow light in other areas and the larger the value of $J$, the more significant the fast light effect. In Fig.\ref{fig:group-speed-with-J}($c$), the system can achieve fast light in the $-\kappa_{1}<\kappa_{2}<-0.6\kappa_{1}$, $0.5\kappa_{1}<J<0.68\kappa_{1}$, and slow light
in other areas. This result shows that, when other parameters are appropriate, we can switch between fast and slow light
by adjusting the distance between the coupled cavities (adjusting $J$). However, when the normalized optical detuning $\delta^\prime=2.316$, we find that only the slow light effect appears in Fig.\ref{fig:group-speed-with-J}($d$) and Fig.\ref{fig:group-speed-with-J}($e$), and the larger the value of $J$, the more obvious the slow
light effect. Different from them, in the Fig.\ref{fig:group-speed-with-J}($f$), one can see that when $\kappa_{2}=-0.7\kappa_{1}$, $J=0.66\kappa_{1}$, ultra-fast light can still be achieved, that is, $\tau_{g}=-1$ms. This
shows that, when there are two different mechanical membranes in the system, $J$ is still
the system optical switch under the current parameters.\\

\begin{figure}[t!]
	\centering\includegraphics[width=0.5\linewidth]{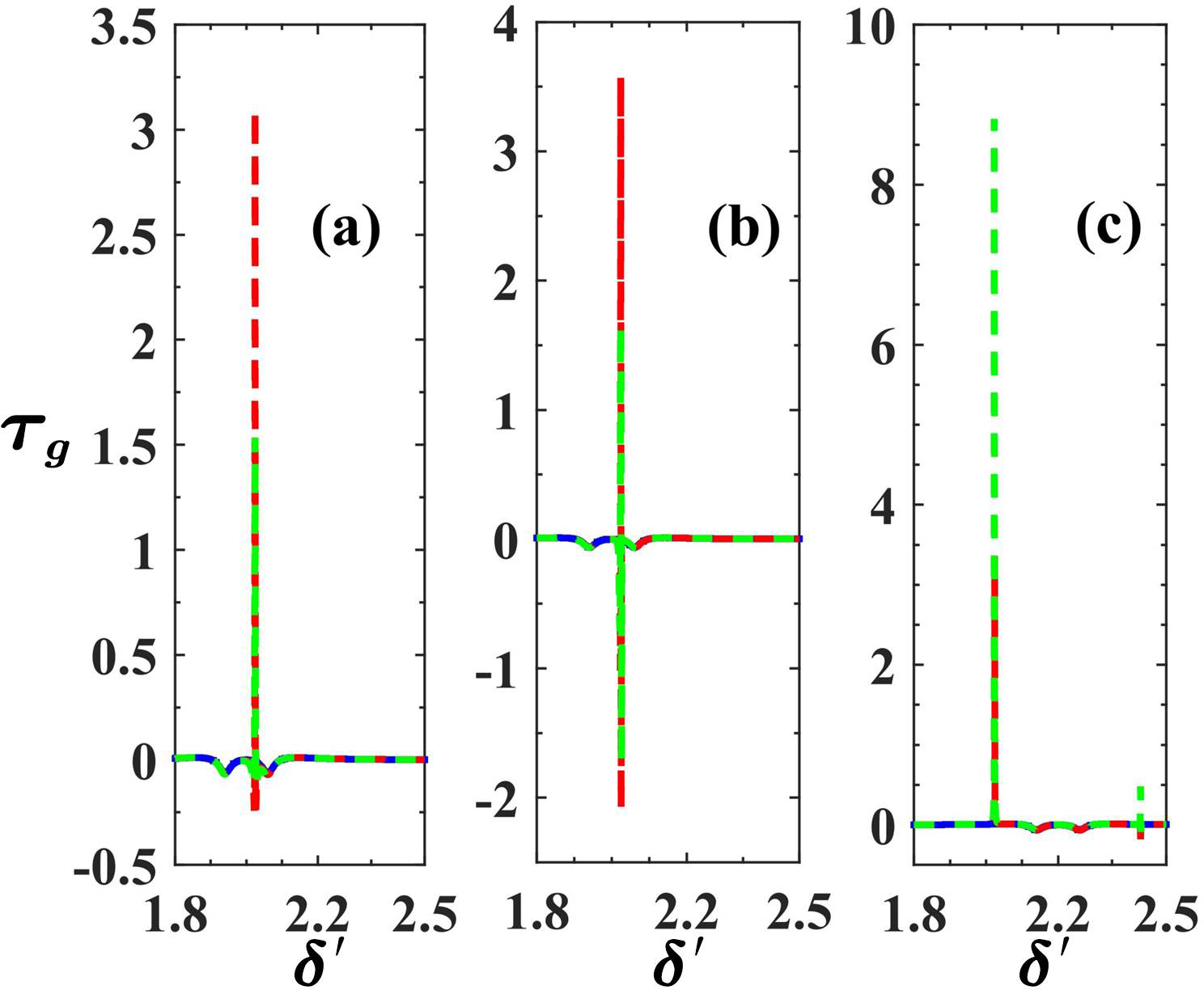}
	\caption{\small The group speed $\tau_{g}$ as functions of the normalized optical detuning $\delta^\prime$ for different environment temperature, where the Figs.14($a$), 14($b$) and 14($c$) represent only one mechanical
		membrane ($g_2=0$), two identical mechanical membranes ($\omega_{m1}=\omega_{m2}$, $\omega_{b1}=\omega_{b2}$), two different mechanical membranes ($\omega_{m2}=1.2\omega_{m1}$, $\omega_{b2}=1.2\omega_{b1}$), respectively . The blue, red and green curves indicate that $T=0$K, $40$K and $90$K, respectively.}
	\label{fig:group-speed-with-normalized-detuning-diff-temp}
\end{figure}
As shown in Fig.\ref{fig:group-speed-with-normalized-detuning-diff-temp}, the group speed $\tau_{g}$ (in ms) as functions of the normalized optical detuning $\delta^\prime$ for different environment temperature. According to the Fig.\ref{fig:group-speed-with-normalized-detuning-diff-temp}, we can find that the temperature can adjust the strength of the fast and slow light effect near the transparent window caused by the two-phonon process in the quadratic coupling optomechanical system, but it has little effect on the fast and slow light effects near the transparent window caused by photon-tunneling. This result is consistent with the previous study on the influence of temperature on the OMIT window. In particular, even under high temperature conditions, the system we studied can still achieve ultra-slow light. However, different from effective pump rate of active cavity $\kappa_{2}$ and photon-tunneling strength $J$, one can see that environment temperature can only adjust the strength of the fast and slow light effect, but cannot switch between fast and slow light.

\section{Conclusion}\label{sec:sec4}
As a summary of our article, we theoretically investigated a dual membrane active-passive
	cavity where each cavity mode is quadratically interacting with displacement
	of the mechanical membrane. Using Keldysh Green’s functional approach we first
	calculated the full retarded Green’s function, then the transmission coefficient of the
	system is obtained from this result. We have evaluated the change of the transmission
	coefficient with the photon-tunneling, the effective pump rate, the environment
	temperature as well as damping of the mechanical membrane. Furthermore, we have
	studied the fast and slow light phenomena in this active-passive cavity optomechanical
	system.
	The main results obtained in this work include: (i) the optical response is determined
	by the competition between the photon tunneling, two-phonon process, and gain effects;
	(ii) the photon-tunneling adjusts the frequency of the supermode formed by the two
	cavity modes, thereby adjusting the width and depth of the transparent window; (iii)
	when there is only one mechanical membrane or two identical mechanical membranes
	in the system, with the increase of the effective pump rate, the transparent window
	caused by the two-phonon process gradually becomes shallower and then converted
	into an absorption peak, and the absorption coefficient can exceed one; (iv) the ambient
	temperature promotes the transparent window caused by the two-phonon process; (v)
	by adjusting the effective pump rate of active cavity or the photon-tunneling strength,
	this system can achieve ultra-fast light/ultra-slow light, but a switch
	between fast and slow light and vice versa has also been realized; (vi) though environment temperature can adjust the
	strength of fast and slow light effect, but it can not realize fast and slow light conversion.
	We believe that our study has potential applications in quantum information processing,
	quantum optical devices as well as in quantum metrology and optical memory.

\end{document}